\documentclass[twocolumn,aps,prx,superscriptaddress, 10pt]{revtex4-2}

\usepackage[english]{babel}
\usepackage{graphicx}
\usepackage{dcolumn}
\usepackage[dvipsnames]{xcolor}
\usepackage[mathlines]{lineno}
\usepackage{braket}
\usepackage{tabularx}
\usepackage{amsmath}
\usepackage{mathtools}
\usepackage[noend]{algpseudocode}
\usepackage{algorithm}
\usepackage{hyperref} 
\usepackage{amsfonts}
\usepackage{amssymb}
\usepackage{dsfont}
\usepackage{booktabs}
\usepackage{subfigure}
\usepackage{subcaption}
\usepackage{csquotes}
\usepackage{ragged2e}
\usepackage{amsthm}
\usepackage{makecell}
\usepackage{dsfont}

\newcommand{\id}[0]{{\mathds{I}}}
\newcommand{\dd}[0]{{\mathrm{d}}}

\newcommand{\tr}[1]{{\text{Tr}\left( #1 \right)}}
\newcommand{\trsquare}[1]{{ \text{Tr} \left[ #1 \right] }}
\newcommand{\trsquareS}[1]{{ \text{Tr}_S \left[ #1 \right] }}
\newcommand{\myexp}[1]{\text{e}^{#1}}

\newcommand{\doubleket}[2]{{\left| #1, #2 \right \rangle}}
\newcommand{\doublebra}[2]{{\left \langle #1, #2 \right|}}
\newcommand{\singleket}[1]{{\left| #1 \right \rangle}}
\newcommand{\singlebra}[1]{{\left \langle #1 \right|}}

\newcommand{\parameter}[0]{{\varphi}}
\newcommand{\generator}[0]{{\hat{G}}}

\newcommand{\fluctuation}[0]{F}
\newcommand{\corrmodes}[0]{\gamma}

\newcommand{\bout}[0]{\hat{b}_{\text{out}}}
\newcommand{\bin}[0]{\hat{b}_{\text{in}}}

\newcommand{\norm}[1]{\left\lVert#1\right\rVert}

\begin{document}
	
\title{Extracting information from a superradiant burst using simple measurements}

\author{Federico Belliardo}
\email {federico.belliardo@gmail.com}
\affiliation{ 
    Pritzker School of Molecular Engineering, University of Chicago, Chicago, Illinois 60637, USA
}

\author{Anjun Chu}
\affiliation{ 
    Pritzker School of Molecular Engineering, University of Chicago, Chicago, Illinois 60637, USA
}

\author{Martin Koppenh\"ofer}
\affiliation{Fraunhofer Institute for Applied Solid State Physics IAF, Tullastr. 72, 79108 Freiburg, Germany}

\author{Aashish A. Clerk}
\email {aaclerk@uchicago.edu}
\affiliation{ 
    Pritzker School of Molecular Engineering, University of Chicago, Chicago, Illinois 60637, USA
}

\begin{abstract}
It is well known that superradiant decay of an ensemble of $N$ spins generates a complex non-classical state of light.  Here, we consider the information content of a superradiant burst of photons: how is information encoded in the initial spin state distributed among the emitted photons, and can it be extracted using simple measurements?  Despite the complexity of the photonic burst state, we show that a simple homodyne measurement combined with an optimized filter and linear estimator recovers the $N$-scaling of the quantum Fisher information of the initial spin state (including cases exhibiting $N^2$ Heisenberg scaling).  Even more surprising, the temporal mode with optimal information content contains a vanishing fraction of the total emitted photons in the large-$N$ limit, suggesting an effective compressing of information.  Our results and setup represent a new way to perform cavity based readout of solid-state spin ensembles that allows one to utilize resonant spin-photon interactions. 
\end{abstract}

\maketitle

\vspace{5mm}

\section{Introduction}
\label{sec:introduction}

Dicke superradiance, the sudden and bright burst of radiation from an ensemble of $N$ coherently emitting atoms, is the perhaps the most paradigmatic example of a collective many-body phenomena in quantum optics~\cite{dicke_coherence_1954, gross_superradiance_1982, rehler_superradiance_1971}.  It has seen a recent resurgence of interest, spurred both by new theoretical developments 
(see, e.g., Ref.~\cite{asenjo-garciaExponentialImprovementPhoton2017,Mivehvar2017,masson_universality_2022,malz_large-n_2022,Rubies2023,Sundar2024,Zhang2025}),
as well as experiments (see, e.g., Ref.~\cite{bohnet_steady-state_2012,Yan2023,kristensenSubnaturalLinewidthSuperradiant2023,Liedl2024,songDissipationinducedSuperradiantTransition2025,kerstenSelfinducedSuperradiantMasing2026}).
There is considerable interest in trying to understand and even harness the properties of the complex light field generated by this emission process.  Recent work has studied how photons in the burst field are non-trivially distributed across multiple temporal modes \cite{van_dorsselaer_collective_1997, perarnau-llobetMultimodeFockStates2020, lemberger_radiation_2021}, and how its non-classical properties could be used to drive an interferometer, enabling quantum-enhanced phase estimation~\cite{gonzalez-tudela_deterministic_2015,paulisch_quantum_2019}. 

\begin{figure}[tbp]
  \centering
  \includegraphics[width=0.45\textwidth]{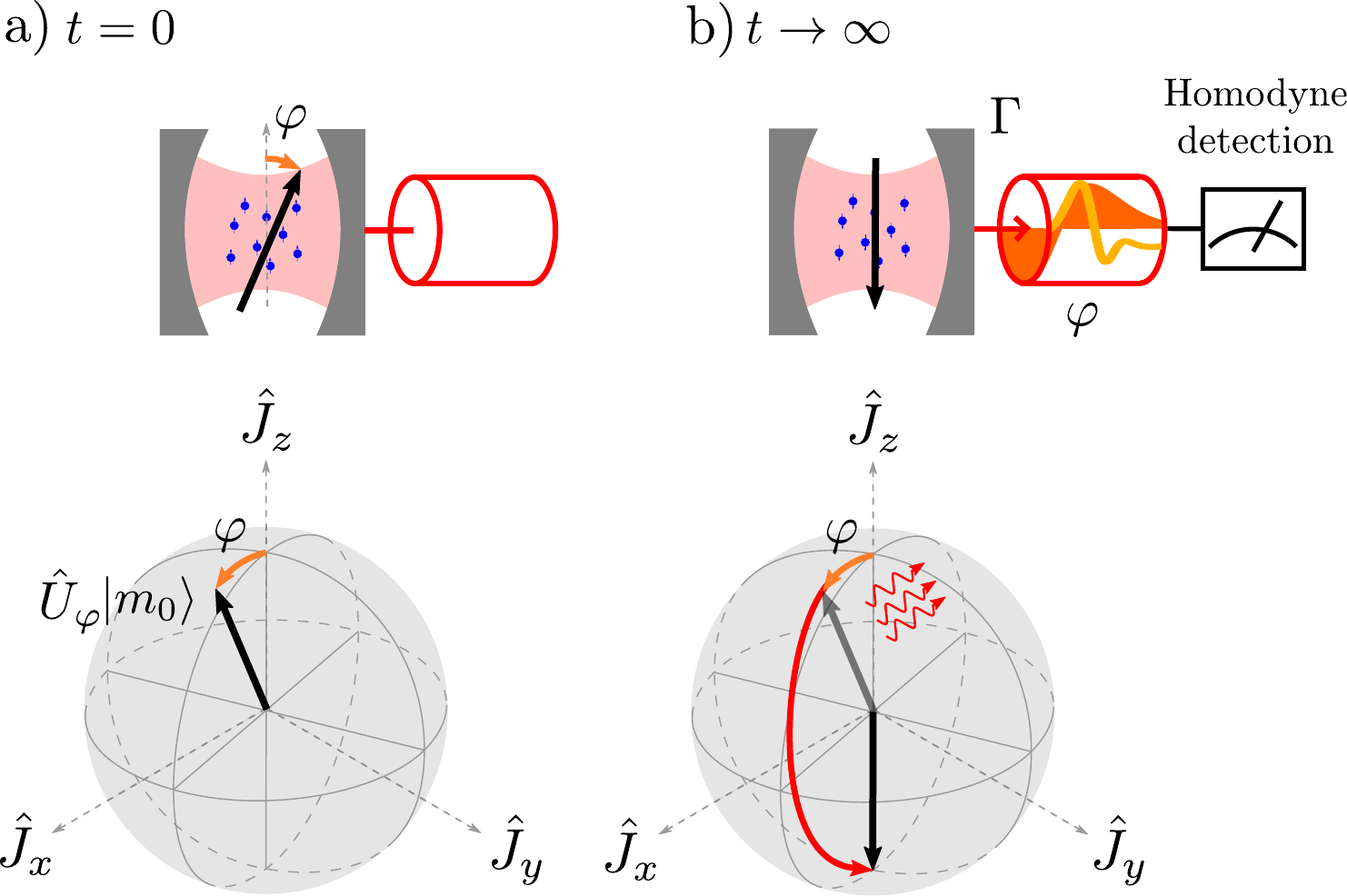} 
  \caption{\justifying a) Generic setup and initial state: spin ensemble (with collective angular momentum operators $\hat{J}_\alpha$) in a cavity, where the spins have been infinitesimally perturbed from the fully-excited state by a parameter-dependent unitary $\hat{U}_\varphi$. In the picture, the rotation angle $\parameter \simeq 0$ has been enhanced for clarity. This initial spin state has a quantum Fisher information $\mathcal{F}_\varphi$.   
  b) Long-time state: the spin ensemble decays superradiantly, with a $\varphi$-dependent photonic burst state generated in multiple temporal modes of the cavity's input-output waveguide.  We show that for a variety of cases, a simple homodyne measurement that targets a single temporal mode of the output field can recover a constant fraction of $\mathcal{F}_\varphi$ in the large-$N$ limit, even in cases where one has Heisenberg scaling.  }  
  \label{fig:stages_metrology}
\end{figure}

In this work, we turn to another aspect of superradiant emission that has both fundamental and practical relevance: what is the {\it information content} of the superradiant burst of photons, and can this information be extracted using simple linear measurements?  We imagine the generic situation sketched in Fig.~\ref{fig:stages_metrology}:  a fully excited spin ensemble in a cavity has an infinitesimal parameter $\varphi$ imprinted upon it (e.g., via a standard Ramsey protocol), before decaying superradiantly through the cavity into a waveguide.  The initial information content of the spin ensemble (as quantified by the quantum Fisher information) must necessarily be transferred to the output $N$-photon waveguide state.  Our question is to ask how this information is distributed between the emitted photons and corresponding temporal modes, and whether a simple measurement can be used to extract it.     

Our work provides a comprehensive study of how information in the initial spin state becomes encoded in the photons of the superradiant burst.  It reveals several surprises.  Despite the non-classical and inherently multi-mode nature of the superradiant emission burst, we show that one can extract almost half of the quantum Fisher information of the initial state using an extremely simple measurement protocol: homodyne measurement that isolates a single temporal mode via an optimized filter, combined with a simple linear estimator.  This result remains true even for cases where the Fisher information exhibits Heisenberg scaling with the number of spins (via use of the so-called twist-untwist protocol~\cite{davis_approaching_2016,hosten_quantum_2016}).  Equally  surprising, we find that the temporal mode with optimal information content is not the same as the mode that has the maximum number of photons.  While for large $N$ it is known that greater than $90\%$ of emitted photons occupy a single temporal mode \cite{lemberger_radiation_2021}, we find that the temporal mode optimizing information in a homodyne measurement contains a vanishing fraction of the emitted photons: in the simplest case, the number of photons in this mode scales only as $\sim N^{2/3}$. For cases where the initial spin state has a quantum Fisher information with SQL scaling, we show that one can extract {\it all} of the information with photon counting or homodyne measurement if one goes beyond a linear estimator and makes use of higher moments.  We show this by making a connection between our setup and the problem of continuous metrology~\cite{albarelli_ultimate_2017, albarelliPedagogicalIntroductionContinuously2024}, and extending the recently derived results of Ref.~\cite{mattes_designing_2025}.   

Note that while the basic question we ask is of fundamental interest, it also has relevance to a variety of leading platforms that are being pursued for quantum metrology applications.  There is growing interest in pursuing quantum metrology using solid state spin ensembles (e.g., NV center defect spins in diamond) coupled to microwave or nanophotonic cavities, as well as studies of superradiance in such platforms (see, e.g., Ref.~\cite{angerer_superradiant_2018,bradac_room-temperature_2017, borkowski_revisiting_2026}).  While the standard approach to cavity-based readout in such cases would be to use a non-resonant dispersive interaction \cite{mamaevDetectingQuantumNoise2026a}, in some cases resonant approaches could be beneficial as they  make use of the full cavity-spin coupling strength, and hence could work considerably faster. Our analysis exactly describes such a resonant protocol where information is resonantly transferred from the spin ensemble to an outgoing photonic or microwave field via a superradiant decay. We also analyze cases where one uses a non-resonant spin-cavity interaction (still far from the dispersive regime) in order to slow down the superradiant burst; even in this case, we show that our basic approach is effective.  A homodyne measurement using an optimal filter is still able to extract almost half of the initial state's quantum Fisher information, despite the increased complexity of the dynamics.

The rest of this manuscript is organized as follows. In Sec.~\ref{sec:quantum_metrology_spin_ensemble} we introduce the physics of the superradiant ensemble of spins and our metrological spin-readout problem. In Sec.~\ref{sec:optimal_modes} we define temporal modes of the output field and apply homodyne detection to the estimation problem. In Sec.~\ref{sec:main_results} we derive the optimal temporal mode to measure in homodyne detection with a linear estimator (i.e.~the mode that maximizes the signal-to-noise ratio), and discuss its properties. In Sec.~\ref{sec:cavity_detuning} we discuss how our results change when including collective spin-spin interactions (arising from a non-zero spin-cavity detuning). Finally in Sec.~\ref{sec:saturation_qfi} we discuss the achievability of the ultimate quantum limit going beyond the use of filters and linear estimators.

Note that our setting is distinct from works studying metrology with SR lasers \cite{bohnet_active_2013, bohnet_steady-state_2012, weiner_superradiant_2012, svidzinsky_quantum_2013}, which focus on continuous estimation of the coherence of a pumped spin ensemble; in contrast, our goal is to read out information initially encoded in an ensemble without the use of any driving.  It also has minimal overlap with other works studying superradiance for metrology (e.g.~non-classical input states for interferometers~\cite{paulisch_quantum_2019, gonzalez-tudela_deterministic_2015, perarnau-llobetMultimodeFockStates2020}, readout using spin amplifiers~\cite{koppenhofer_dissipative_2022, koppenhofer_squeezed_2023}), readout using disordered superradiance \cite{Bohr2024}).

\section{Model and generic quantum metrology protocol with a spin-ensemble sensor}
\label{sec:quantum_metrology_spin_ensemble}

\subsection{Basic model}

We consider a generic setup (as sketched in Fig.~\ref{fig:stages_metrology}) where an ensemble of $N$ identical spin-$1/2$ systems acts a sensor, and is coupled to a single bosonic mode (a cavity or resonator mode) that will ultimately be used for readout.  Working in a rotating frame set by the spin Larmor frequency, the Hamiltonian of the system takes the form of a Tavis-Cummings model:  
\begin{equation}
    \hat{H} = \Delta \hat{a}^\dagger \hat{a} + g \left( \hat{J}_+ \hat{a} + \hat{J}_{-} \hat{a}^\dagger \right) \; ,
    \label{eq:hamiltonian_tavis_cummings}
\end{equation}
where $\Delta$ is the detuning of the cavity and the spins, and $g$ is the strength of the collective coupling of the spins and the cavity. $\hat{a}$ is the cavity mode annihilation operator, while the spin ensemble is described by collective angular momentum raising and lowering operators, $\hat{J}_+$ and  $\hat{J}_-$.  We will use $\singleket{m}$ to denote standard Dicke states, i.e.~states with a maximum total angular momentum $j_{\rm tot} = N/2$ and a $\hat{J}_z$ eigenvalue $m$. We focus throughout on fully collective dynamics that conserves the total angular momentum of the spin ensemble.  

We work in the standard regime where the coupling between the spins and the cavity is sufficiently weak compared to the loss of the cavity (``bad-cavity'' regime), such that the cavity can be adiabatically eliminated~\cite{bonifacio_quantum_1971}. In this limit, the density matrix $\hat \rho$ of the spin ensemble obeys a Lindblad-style master equation that has cavity-mediated collective interactions and collective decay:
\begin{equation}
    \frac{\dd \hat{\rho} (t)}{\dd t} = i \left[ \chi \hat{J}_z^2, \hat{\rho} (t) \right]  + \mathcal{D} [ \sqrt{\Gamma} \hat{J}_{-} ] \hat{\rho} (t)\; ,
    \label{eq:lindbladian_decay}
\end{equation}
with $\Gamma \equiv 4 g^2 \kappa / (4 \Delta^2 + \kappa^2)$ and $\chi \equiv 4 g^2 \Delta / (4 \Delta^2 + \kappa^2)$, and $\mathcal{D}[\sqrt{\Gamma} \hat{J}_{-}] \left( \cdot \right) =  \Gamma \left( \hat{J}_{-} \cdot \hat{J}_{+} - \frac{1}{2} \lbrace \hat{J}_+ \hat{J}_-, \cdot \rbrace \right)$. 
Note that Eq.~(\ref{eq:lindbladian_decay}) has already been realized experimentally (see, e.g., Ref.~\cite{norcia2018}). 

\subsection{Encoding information in the initial spin state}

The basic sensing protocol we study corresponds to a local parameter estimation problem, and is
depicted schematically in Fig.~\ref{fig:stages_metrology}:  a fully excited spin ensemble has an infinitesimal parameter $\parameter$ encoded in it, after which it decays superradiantly, with information transferred to the photonic state (collected in a waveguide).  Our goal is to understand how information on $\parameter$ is distributed between the photons and temporal modes of the emitted photonic field, and whether it can be simply extracted using a linear homodyne measurement.


We start our analysis by discussing the first information-imprinting step in the protocol.
We assume that the initial state of the spin ensemble $\ket{\psi_\parameter}$ is given by the action of an infinitesimal unitary $\hat{U}_\parameter$ acting on a fully-excited product state $\ket{m=m_0}$ with $m_0 = N/2$:
\begin{equation}
    \ket{\psi_\parameter} \equiv \hat{U}_\parameter  \singleket{m_0} \, ,  \; \text{with} \; \hat{U}_\parameter \equiv \myexp{i \parameter \generator} \; .
    \label{eq:generic_initial_state}
\end{equation}
We restrict our attention to cases where the parameter generator $\generator$ is also a collective spin operator (i.e., it commutes with $\hat{J}^2$), e.g., $\hat{G} = \hat{J}_y$. The information content of the initial spin state is quantified by the quantum Fisher information (QFI), given by:
\begin{equation}
    \mathcal{F}_\parameter(\ket{\psi_\parameter}) = 4 \left( \braket{m_0| \generator^2 |m_0} - \braket{m_0| \generator |m_0}^2 \right) \; .
    \label{eq:standard_quantum_limit}
\end{equation}
%
The most celebrated result of quantum metrology is the possibility of achieving Heisenberg scaling (HS)~\cite{giovannetti_advances_2011}, i.e., $\mathcal{F}_\parameter(\ket{\psi_\parameter}) \propto N^2$, by using an entangled state of $N$ probes (e.g., spins), instead of $\mathcal{F}_\parameter(\ket{\psi_\parameter}) \propto N$, which is the standard quantum limit (SQL) and arises when a separable state is used~\cite{giovannetti_advances_2011, pezze_quantum_2018}. Note that it necessarily follows that at long times (after the superradiant decay is over), the QFI of the photonic state will be identical to the QFI of the initial spin states.  

In what follows, we will focus on two specific, experimentally-relevant choices for $\hat{G}$. The first corresponds to $\hat{U}_\parameter$ performing a rotation of the spin ensemble around the $y$-axis, i.e., $\generator = \hat{J}_y$. In this scenario the encoded state is
\begin{equation}
    \ket{\psi_\parameter^S} \equiv \hat{U}_\parameter  \singleket{m_0} \, ,  \; \text{with} \; \hat{U}_\parameter \equiv \myexp{i \parameter \hat{J}_y} \; ,
    \label{eq:initial_state_separable}
\end{equation}
which is separable and has $\mathcal{F}_\parameter(\ket{\psi_\parameter^S})=N$. 
We stress that the spin coherent state $\ket{\psi_\parameter^S}$ is exactly the kind of state generated by a standard Ramsey sensing protocol \cite{degen_quantum_2017}; the only difference here is that a final rotation is performed so that the spin ensemble is almost perfectly polarized in the $+z$ direction.  

The second special generator we consider originates from the celebrated twist-untwist strategy for Heisenberg-limited metrology~\cite{hosten_quantum_2016, davis_approaching_2016, scharnagl_optimal_2023, koppenhofer_revisiting_2023}. This strategy makes use of squeezing and anti-squeezing, and is known to allow one to reach HS with simple (even imperfect) projective measurements of the spin ensemble.  We show that it is also useful when combined with the superradiant readout strategy sketched in Fig.~\ref{fig:stages_metrology}.  The twist-untwist protocol makes use of a unitary squeezing operator  $\hat{U}_{s} \equiv \myexp{i Q \hat{J}_x^2/N}$ generated by a one-axis twisting Hamiltonian.  It involves three stages: i) squeeze an initial product state, ii) perform a $\parameter$-dependent rotation of the spin ensemble, e.g., via a Ramsey sequence, iii) perform an anti-squeezing operator corresponding to the unitary $\hat{U}_s^\dagger$.  The resulting $\parameter$-dependent state (upon suitable rotation) thus takes the form
\begin{equation}
    \ket{\psi_\parameter^E} \equiv \myexp{i \parameter \, \generator}  \singleket{m_0} \, ,  \; \text{with} \; \generator = \hat{U}_s^\dagger \hat{J}_y \hat{U}_s \; ,
\label{eq:initial_state_entangled}
\end{equation}
This state has the same QFI as the entangled state $\hat{U}_s |m_0 \rangle$ (with respect to a parameter generator $\hat{J}_y$).  Note that as a function of the squeezing parameter $Q$, the signal-to-noise ratio (SNR) of a projective measurement of  $\hat{J}_y$ (corresponding to a simple linear estimator) is maximized at $Q = Q_{\rm opt} \equiv N \text{arccot} (N - 2)$ \cite{davis_approaching_2016}.  For this choice of $Q$, both the SNR and QFI exhibit Heisenberg scaling with $N$ (although with different prefactors, see App.~\ref{sec:ideal_linear_measurement}).  As our homodyne strategy seeks to replicate the information available in a direct $\hat{J}_y$ measurement, we will also set $Q = Q_{\rm opt}$ in our analysis.

\section{Photonic output field: temporal modes and homodyne measurement}
\label{sec:optimal_modes}

We return to Fig.~\ref{fig:stages_metrology}, and now discuss the readout aspect of our generic metrology protocol, where the superradiant decay of the spin ensembles creates a propagating photonic state (the superradiant burst) whose quantum state depends on the parameter $\parameter$ that was originally encoded in the spin ensemble.
Our goal is to understand the information content of these photons, and whether a simple linear homodyne measurement is sufficient to capture this information.  To set the stage for this discussion, we review homodyne measurement in our setup, as well as the temporal mode description of the output field.  

\subsection{Input-output theory and homodyne measurement}

The photonic output field in the cavity is described by the operator $\hat{b}_{\rm out}(t)$, as determined by standard quantum-optical input-output theory~\cite{clerkIntroductionQuantumNoise2010, gardinerQuantumNoiseHandbook2004}.  As we are working in the regime where the cavity  mode $\hat{a}$ has been adiabatically eliminated, the basic input-output equation for the setup in Eq.(\ref{eq:lindbladian_decay}) takes the form:
\begin{equation}
    \bout(t) =  -i \sqrt{\Gamma} \hat{J_{-}}(t) + \bin (t) \; .
    \label{eq:input_output_theory}
\end{equation}
Both input and output fields satisfy $ [\hat{b}_\alpha (t), \hat{b}_\alpha^\dagger (t')]=\delta(t-t')$, with $\alpha = \text{out}/\text{in}$, and the input field is taken to be vacuum noise (there is no driving of the cavity), implying  $\langle \bin(t) \bin^\dagger(t') \rangle = 0$.  

At least some information on the parameter $\varphi$ will be available in $\langle \bout(t) \rangle$ for times $t>0$.  We thus define the {\it signal function} $\alpha_I(t)$ by the parametric derivative of this average value with respect to $\parameter$:
\begin{equation}
    \alpha_I(t) \equiv 
    \frac{\partial \langle \bout(t) \rangle_{\parameter}}{\partial \parameter} \bigg|_{\parameter = 0} \; .
\end{equation}
Using the input-output relation in Eq.~(\ref{eq:input_output_theory}) one finds that the signal function is given by:
\begin{align}
     \alpha_I(t) & = 
    -i \sqrt{\Gamma} \frac{\partial \langle \hat{J}_{-} (t) \rangle_{\parameter}}{\partial \parameter} \bigg|_{\parameter = 0} =  \sqrt{\Gamma} \, \langle m_0| [  \hat{J}_{-}(t),\generator ] | m_0 \rangle \; .
\label{eq:homodyne_sensitivity}
\end{align}
This relation directly follows from the definition of the average, and reflects the structure of Kubo linear-response susceptibility (i.e., how $\hat{J}_-$ responds to an initial perturbation involving $\hat{G}$). In App.~\ref{sec:analytical_signal}, we compute the signal $\alpha_I(t)$ for the superradiant decay described by Eq.~\eqref{eq:lindbladian_decay}.

One might expect that the full time-dependence of $\alpha_I(t)$ would have a strong sensitivity to the form of $\hat{G}$.  Surprisingly, this is not the case.  The dynamics in Eq.~\eqref{eq:lindbladian_decay} has a weak $U(1)$ symmetry generated by $\hat{J}_z$~\cite{albertSymmetriesConservedQuantities2014}, which gives the Lindbladian a block-diagonal structure, greatly constraining dynamics.  A consequence is that for any collective generator $\generator$, we have (see App.~\ref{app:separation_theorem}):
\begin{align}
    \alpha_I(t) &= \singlebra{m_0}\generator\singleket{m_0-1} \cdot  \sqrt{\Gamma} \braket{m_0-1 | \hat{J}_{-}(t) | m_0} \\ &\equiv 
    \singlebra{m_0}\generator\singleket{m_0-1} \cdot \tilde{\alpha}_I (t) \; ,
\label{eq:universal_expression_signal}
\end{align}
We see that the time dependence of the signal function has a ``universal" dependence on time, in that its shape does depend on the choice of generator. The generator only sets an overall scale via the prefactor.  This results continues to hold for the full collective dynamics described by Eq.\eqref{eq:lindbladian_decay} (which has collective decay as well as spin-spin interactions), and even holds if we do not adiabatically eliminate the cavity and work with the cavity in Eq.~\eqref{eq:hamiltonian_tavis_cummings}, with whatever collective spin-spin interaction.

The information in the average output field can be captured by a homodyne measurement, a simple and ubiquitous linear measurement that yields a current $I(t)$ that is proportional to one quadrature of the output field.  Using a time-dependent homodyne angle $\theta(t)$, the homodyne current operator is:
\begin{equation}
    \hat{I}(t) \equiv e^{i \theta (t)} \bout(t) + \text{h.c.} \; .
    \label{eq:homodyne_current}
\end{equation}
If one takes $\theta(t) = -\arg \alpha_I(t)$, then we have that the average homodyne current attains its maximum value, and $\partial_\varphi \langle \hat I(t) \rangle = 2 | \alpha_I(t)|$. Homodyne measurements are standard in both quantum optical setups (where they are realized using local oscillators, beam-splitters, and photocounting~\cite{wallsQuantumOptics2024}), as well as in microwave circuit QED setups, where they can be directly implemented using an IQ mixer. 

To analyze the information available from a homodyne measurement, we also need to understand the fluctuations of $\hat{I}(t)$.  As we are always considering an infinitesimal parameter $\parameter$, we can calculate these fluctuations at $\parameter=0$.  Further, our main focus will be to use a simple linear estimator for $\parameter$, i.e., the estimate of $\parameter$ will be proportional to the (optimally weighted) time-integrated homodyne current. See App.~\ref{sec:definition_information_quantities} for the definition of the estimator and its precision. The estimation error will then be determined by a signal-to-noise ratio SNR, and we only need to understand the homodyne current fluctuations at the level of their auto-correlation function (evaluated at $\parameter = 0$):
%
\begin{align}
    S_{II} (t', t) &\equiv \langle \hat{I}(t') \hat{I} (t) \rangle - \langle \hat{I}(t') \rangle \langle \hat{I}(t) \rangle \; . \label{eq:original_S}
\end{align}
%


\subsection{Temporal modes}

Our goal will be to find an optimal filter for the homodyne current, in such a way that maximizes the resulting SNR.  This filtering naturally corresponds to a temporal mode decomposition of the output field.  We briefly review this notion here.  We introduce a mode function $u(t)$ (chosen to have unit ${L}_2[0, \infty]$ norm),
and an associated bosonic operator
\begin{equation}
    \hat{b}_u \equiv \int_{0}^{\infty} u(t) \bout (t) \dd t \;.
\end{equation}
This is a canonical bosonic annihilation operator satisfying $[\hat{b}_u, \hat{b}_u^\dagger] = 1$; it corresponds to removing photons from the chosen temporal mode.  
By picking a set of mode functions that form a complete orthogonal basis, we can describe the output field using a set of commuting temporal mode operators $\lbrace \hat{b}_{u_j} \rbrace_{j=1}^\infty$. This gives a decomposition of the output field in a discrete basis of wavepackets~\cite{clerkIntroductionQuantumNoise2010}.





There is a long history of analyzing the output field produced by superradiant decay in terms of temporal modes and their correlations (see, e.g., \cite{law_dynamic_2007,lemberger_radiation_2021}).  The standard focus has been to find an orthogonal mode basis where there are no first-order coherences between the modes.  This corresponds to diagonalizing the $g_1(t,t')$ correlation matrix:
\begin{equation}
    g_1(t', t) \equiv \Gamma \langle m_0 | \hat{J}_{+} (t') \hat{J}_{-}(t) | m_0 \rangle \; .
    \label{eq:g_1_t_t}
\end{equation}

\begin{figure*}[tbp]
  \centering
  \includegraphics[width=\textwidth]{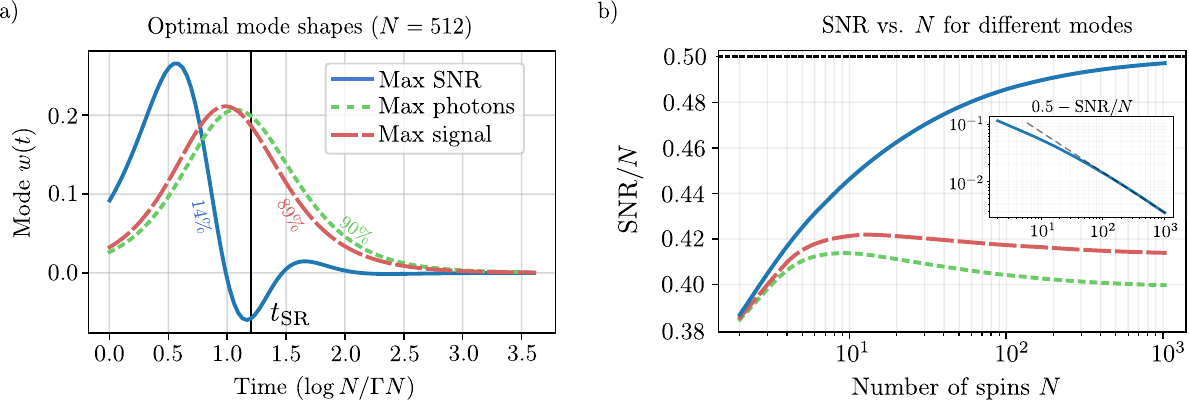} 
  \caption{\justifying a) Mode profiles $w(t)$ for the maximum-photons mode (Eq.~\eqref{eq:most_populated_mode}), the maximum-signal mode (Eq.~\eqref{eq:most_sensitive}) and the maximum-SNR mode (Eq.~\eqref{eq:snr_optimizer}).  Results are shown for pure superradiant decay dynamics (i.e., $\chi = 0$ in Eq.~(\ref{eq:lindbladian_decay})), in which case these modes are real and independent of the generator $\generator$ (up to an overall time-independent phase). We take $N=512$. Each curve is labeled by the fractional photon number occupation of each mode $\bar{p}_u = \bar{n}_u / N$; for the maximum-SNR mode this ratio vanishes in the large-$N$ limit. The vertical line indicates $t=t_\text{SR} \simeq \frac{\log N}{\Gamma N}$, the time associated with the peak intensity of the superradiant burst. b) For a simple Ramsey encoding of a rotation angle, as described in Eq.~\eqref{eq:initial_state_separable}, we evaluate the ratio of the SNR to the quantum Fisher information (which has the value $N$) as a function of $N$.  For each mode shape this ratio converges to a constant as $N \rightarrow \infty$; in particular, for the maximum-SNR mode (Eq.~\eqref{eq:snr_optimizer}), the ratio converges to $0.5$. 
  The inset shows convergence to this value as a power law, $\frac{1}{2} - \frac{\text{SNR}}{N} \simeq 0.30 \cdot N^{-0.66}$. The linestyles and linecolors label the three modes the same way as a). The maximum-photons and maximum-signal modes converge to a SNR lower than the SNR of $w_\text{SNR}(t)$, as expected.}
  \label{fig:three_modes_comparison}
\end{figure*}

Given some mode $u(t)$, the most basic question is the computation of its average photon number $\bar n_u$:
\begin{equation}
    \bar n_u \equiv \langle \hat{b}^\dagger_u \hat{b}_u \rangle =  \int_{0}^\infty \dd t' \dd t \, u^{*} (t') g_1 (t', t) u(t) \; .
    \label{eq:photon_number}
\end{equation}
%
Of particular interest is the temporal mode $w_{\text{pop}}(t)$ that has a maximum average photon number:
\begin{equation}
    w_{\text{pop}}(t) \equiv \operatorname*{argmax}_{\substack{u(t)}} \int_{0}^\infty \dd t' \dd t \, u^{*} (t') g_1(t', t) u(t) \; .
    \label{eq:most_populated_mode}
\end{equation}
This mode is the eigenvector $g_1(t', t)$ with the largest eigenvalue.  The photon number distribution between temporal modes has been extensively studied in the case of pure superradiant decay, i.e., the resonant-cavity limit of Eq.~(\ref{eq:lindbladian_decay}) where $\chi = 0$ and there is only collective decay.  
As discussed in Ref.~\cite{lemberger_radiation_2021} the radiation emitted in superradiance is multimodal, but for large $N$, the most populated mode carries $\sim 90\%$ of the photons, and the fraction of the photons carried by the other orthogonal modes decreases rapidly with the index of the mode. In Fig.~\eqref{fig:three_modes_comparison} we plot $w_{\text{pop}}(t)$ for a superradiant ensemble of spins with $N=512$. 



\subsection{Filtered photodetection has no information}
While the total average number of photons emitted in the superradiant process is fixed, $\bar{n}_{tot} = N$, the distribution of photons across different modes changes with the initial state. Naively, it is possible to think that the distribution of photons across different modes carries information about the parameters $\parameter$, e.g., photon counting on a single mode would allow one to construct a precise estimator for $\parameter$. In App.~\ref{sec:photon_number_sensitivity}, we prove that this picture is wrong, and that the leading order in the variation of this distribution is $\mathcal{O}(\parameter^2)$, so that there is no linear estimation that can retrieve the parameter. This result is valid because of the $U(1)$ symmetry of the dynamics and the initial state $\ket{m_0} = \ket{j}$, and applies to whatever hermitian generator $\hat{G}$. We can't exclude that the noise associated to the observation of the average number of photons is also null (we don't compute the noise), so that the measurement is technically informative, being in a zero-signal over zero-noise scenario. This case is however not practical in experiments and won't be discussed further. This picture changes where more complex estimators are considered, or if the time of arrival of the photons is recorded. Nevertheless, this negative result for a linear estimation is a strong motivation to consider to homodyne detection as a possible solution to the initial-value problem. Again we can't exclude that the achievability of the QFI is due to a zero-signal over zero-measurement scenario.

\section{Main results: information content of an optimal homodyne measurement}
\label{sec:main_results}

Having established the basic description of our system, the sensing protocol, and the photonic output field, we now turn to our central question: {\it how much information on $\parameter$ can be extracted from an optimized homodyne measurement}?  We further restrict attention to the simplest and most experimentally friendly scenario, where the estimator for $\parameter$ is proportional to some weighted time integral of the measured homodyne current.  The question then becomes one of filtering: how do we filter the homodyne current to optimize the SNR of the measurement?
This is equivalent to finding the optimal temporal mode $u(t)$ whose associated integrated homodyne current maximizes the SNR.  A natural guess is that, since photons carry information on $\parameter$, the more photons the better, hence one should use the same temporal mode that maximizes the average photon number, 
c.f.~Eq.~\eqref{eq:most_populated_mode}. Surprisingly, we now show that this guess is incorrect.  In this section, we focus on the case of a resonant cavity, where $\chi = 0$ in the dynamics described by Eq.~(\ref{eq:lindbladian_decay}) (i.e., there is just collective superradiant decay, and no additional Hamiltonian spin-spin interactions).  We address the impact of such interactions in the next section.  

\subsection{Optimal modes for maximum signal and maximum SNR}

We first define the filtered homodyne current for a specific temporal mode $u(t)$ 
to be
\begin{equation}
    \hat{I}_u(t) \equiv u(t) \bout(t)+ \text{h.c.} \; ,
    \label{eq:integrated_current_u}
\end{equation}
where the homodyne phase (i.e.,~local-oscillator phase) $\theta(t)$ has been absorbed into the definition of the complex mode function $u(t)$.  
Given a choice of mode function, the measured quantity will be the integrated and filtered homodyne current
\begin{equation}
    \hat{M}_u \equiv \int_0^\infty \dd t \, \hat{I}_u(t) \; .
    \label{eq:observable}
\end{equation}
Our estimate of $\parameter$ will be proportional to the measured value of $\hat{M}_u$, hence we define the net signal associated with this mode to be  
\begin{equation}
    \bar{S}_u \equiv
    \partial_\varphi  \langle \hat{M}_u \rangle =
    2 \operatorname{Re} \int_{0}^\infty u(t) \alpha_I(t) \; .
    \label{eq:derivative_integrated_current}
\end{equation}

We can now easily identify the temporal mode that maximizes the net signal $\bar{S}_u$.  Denoting this mode as $w_{\rm sig}(t)$, we have:
\begin{align}
    w_{\text{sig}} (t) &\equiv \operatorname*{argmax}_{\substack{u(t)}}\,  2 \operatorname{Re} \int_0^\infty \dd t \, u(t) \alpha_I(t) 
    \nonumber \\ &= \frac{\alpha_I^{*}(t)}{\sqrt{\int_0^t \dd t \,|\alpha_I(t)|^2 }} \; .
    \label{eq:most_sensitive}
\end{align}
Not surprisingly, the maximum-signal mode has the same shape and phase as $\alpha_I(t)$, and the mode-optimized value of the signal $\bar{S}_u$ is simply the $L_2$-norm of $\alpha_I(t)$. 


While identifying the maximum-signal mode is straightforward and perhaps somewhat obvious, our goal is more involved: we want the mode $u(t)$ that optimizes the signal-to-noise ratio associated with a measurement of 
$\hat{M}_u$. For a given mode function $u(t)$, the power SNR is defined as:
\begin{equation}
     \operatorname{SNR}[u(t)]  \equiv
        \frac{ | \bar{S}_u |^2}{{\left \langle 
        \left ( 
            \hat{M}_u - \langle \hat{M}_u \rangle \right)^2 \right \rangle }},\label{eq:definition_SNR_M}
\end{equation}
where both the denominator and numerator are evaluated at $\parameter=0$. The numerator is the square of the signal, while the denominator is the variance of the observable, i.e $\text{Var}(\hat{M}_u)$.

In App.~\ref{sec:optimization_snr}, we show how the fluctuations of the observable $\hat{M}_u$ can be computed from the correlation function for the filtered current, i.e.
\begin{equation}
    S_{II, u} (t', t) \equiv \langle \hat{I}_u(t') \hat{I}_u (t) \rangle - \langle \hat{I}_u(t') \rangle \langle \hat{I}_u(t) \rangle \; 
    \label{eq:S_II_mode_u}
\end{equation}
defined in direct analogy with Eq.~\eqref{eq:original_S}. The $U(1)$ weak symmetry of our dynamics under $\hat{J}_z$ rotations implies that there are no squeezing correlations in the output field. For the computation of the fluctuations of a temporal mode, using these symmetries, and Eq.~\eqref{eq:input_output_theory}, we can introduce the fluctuation matrix
\begin{align}
    \fluctuation (t', t) 
   &\equiv 2 \Gamma \langle m_0 | \hat{J}_{+} (t') \hat{J}_{-} (t) | m_0 \rangle + \delta (t' - t)\label{eq:S_spins}\; .
\end{align}
The first term represents the contribution of the spin dynamics to the homodyne noise, whereas the second term is the usual shot-noise floor (stemming from vacuum fluctuations).  

We can now define the temporal mode that maximizes the SNR, $w_{\text{SNR}}(t)$, as:
\begin{align}
    w_{\text{SNR}}(t)
    &\equiv \operatorname*{argmax}_{u(t)}
    \frac{
        4 \big| \operatorname{Re} \int_{0}^{\infty} \! \mathrm{d}t \, u(t)\,\alpha_I(t) \big|^2
    }{
        \int_{0}^{\infty} \! \mathrm{d}t' \int_{0}^{\infty} \! \mathrm{d}t \,
        u^{*}(t)\, \fluctuation (t,t')\, u(t')
    } \;.
    \label{eq:optimal_mode}
\end{align}
In App.~\ref{sec:optimization_snr}, we perform analytically the optimization and find
\begin{equation}
    w_\text{SNR} (t) = \frac{\int_{0}^\infty \dd t' \, \fluctuation^{-1}(t, t') \alpha_I^{*} (t')}{\sqrt{\int_{0}^\infty \dd t \int_{0}^\infty \dd t' \, \alpha_I (t) \fluctuation^{-2}(t, t') \alpha_I^{*} (t')}}\;.
    \label{eq:snr_optimizer}
\end{equation}
Note crucially that the maximum SNR mode $w_\text{SNR} (t)$ is in general not identical to the maximum-signal mode $w_\text{sig} (t)$.  The only exception is the simple case where the homodyne noise is white noise, i.e., $\fluctuation(t,t') \propto \delta(t-t')$.  
For the more general case of temporally correlated noise, the maximum-SNR mode will be a tradeoff between optimizing the integrated signal while at the same time minimizing noise by making use of correlations~\footnote{Note that one does not have to worry about $S_{II}(t,t')$ being non-invertible.  Non-invertibility would imply the existence of one or more modes that have zero fluctuations in one quadrature.  This would in turn imply infinite squeezing, something that is impossible with a finite number of photons.  In our case, there is even a stronger constraint, as the superradiant decay dynamics never generates squeezing correlations in the output field.  }.

For our superradiant decay dynamics, we have an even more striking result.  
By using the factorized expression for $\alpha_I(t)$ given in Eq.~\eqref{eq:universal_expression_signal}, we see that the form of the maximum-SNR model $w_\text{SNR} (t)$ is universal, i.e., it is independent of the specific form of the parameter generator $\generator$ (up to an overall constant phase factor)
We stress that while our derivation is fully quantum mechanical, the final result in Eq.~\eqref{eq:snr_optimizer} matches a well-known result in classical signal processing:  it corresponds to the optimal ``generalized matched filter" for extracting a signal waveform from a background of Gaussian noise (see, e.g., \cite{woodwardProbabilityInformationTheory1953, olshevsky_matched_2005, picinbono_generalized_1990, turin_introduction_1960}). 

We also stress another nice feature of our optimal-mode result:  for the resonant cavity setup (c.f.~Eq.~(\ref{eq:lindbladian_decay})), one finds that $\alpha_I(t) e^{-i \nu}$ is real at all times, where $\nu$ is a time-independent phase determined by the choice of generator $\hat{G}$ 
(i.e., $\nu = \arg \langle m_0 | \hat{G} |m_0-1\rangle$, c.f.~Eq.~(\ref{eq:universal_expression_signal})). As a result, the phase of the maximum SNR mode is constant (modulo $\pi$) for all times.
This yields a significant experimental simplification: the optimal information mode involves measuring the {\it same} output field quadrature at all times, and does not require time-dependent control of the local-oscillator phase $\theta(t)$ in Eq.~(\ref{eq:homodyne_current}).

Using our knowledge of the maximum-SNR mode, we can also directly compute the optimized value of the SNR, i.e., the maximum SNR possible with a linear estimator obtained from the integrated homodyne measurement of a single mode.  We find:
\begin{align}
    \text{SNR} & \bigl[ w_\text{SNR} (t) \bigr] 
    = 4 \int_0^\infty \! \! \dd t \int_0^\infty \! \! \dd t' \,
       \alpha_I (t)\, \fluctuation^{-1}(t, t')\, \alpha_I^{*} (t') 
    \label{eq:optimal_snr_expression} \\[4pt]
    &= \bigl| \braket{m_0 \lvert \generator \rvert m_0-1} \bigr|^2 \nonumber \\[2pt]
    &\quad \times 4 \int_0^\infty \dd t \int_0^\infty \dd t' \,
       \tilde{\alpha}_I (t)\, \fluctuation^{-1}(t, t')\, \tilde{\alpha}_I^{*} (t') \; .
    \label{eq:factorization_snr}
\end{align}

Fig.~\ref{fig:three_modes_comparison} shows plots of the three different ``optimized" modes we have discussed here (optimized either for maximum photon number, maximum net signal, or maximum SNR), for a large value of $N$. Several points are in order. First note that the maximum-signal and maximum-photon modes have a similar shape: they exhibit no nodes, and have a single peak at a time close to the time associated with the superradiant burst.  While the mode shapes are similar, they are not identical.  Second, and more importantly, the maximum-SNR mode has a shape {\it completely different} from the max-photon number mode.  It changes sign and has a node slightly before the time of the superradiant burst. This sign change is a direct consequence of temporally-correlated photonic noise produced by the superradiant burst:  the maximum SNR mode is creating destructive interference from these correlations by having its sign change. In App.~\ref{sec:cascaded_quantum_state}, we compute explicitly the Fisher information and quantum Fisher information of the quantum state of the maximum SNR mode, and we show that this quantities are only slightly above the SNR of the linear measurement reported in this plot.  

\begin{figure}[tbp]
  \centering
  \includegraphics[width=0.45\textwidth]{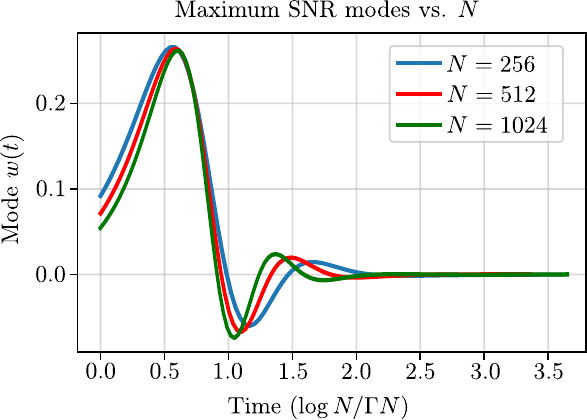} 
  \caption{\justifying Temporal modes maximizing the SNR, as computed in Eq.~\eqref{eq:snr_optimizer}, for different values of the spin number $N$. We do not see any sign of convergence for the shape of the modes, when rescaled for the timescale of the superradiant.}
\label{fig:figure_6}
\end{figure}

While it was suggested numerically in~\cite{lemberger_radiation_2021} that the asymptotic shape of the max-photon mode 
(and all other eigenfunctions of $g_1(t', t)$ defining a canonical orthonormal basis) converges in the large-$N$ limit when the time is appropriately rescaled to the superradiance timescale, we found no evidence for such convergence for the mode maximizing the SNR (see Fig.~\ref{fig:figure_6}). This is consistent with the fact that the number of photons in the optimal mode scales sublinearly with $N$. If an asymptotic shape existed, there would be a stable asymptotic decomposition onto the eigenfunctions of $g_1(t', t)$, implying an asymptotic value for the fractional occupation number, which is in contradiction with the scaling law found in Fig.~\ref{fig:three_modes_comparison}. Since the number of photons in the higher-order modes of the canonical basis decreases exponentially fast with the mode index, while the fractional occupation number only scales as a power law in $N$, these higher-order modes, with increasing number of nodes each, emerge very slowly in $w_{\text{SNR}}(t)$ as $N$ increases. This explains the slowly varying shape of the modes in Fig.~\ref{fig:figure_6}. The most striking observation from this picture is that while the main maximum of the modes seems to have converged, the anti-max and the second max do not seem to be converging, indicating that they might be controlled by another timescale other that the other than what controls the burst.

We stress that the expressions in Eqs.~\eqref{eq:most_populated_mode}, \eqref{eq:most_sensitive}, and \eqref{eq:optimal_mode} are completely general and could be used to analyze the information content of a homodyne measurement of a photonic field produced by arbitrary dynamics (not just the superradiant decay dynamics studied here).  


\subsection{Optimal homodyne measurement with a linear estimator allows Heisenberg-limited scaling}

We are now in a position to state the main key result of our paper (again focusing to start on the resonant-cavity, $\chi = 0$ case): once we optimize the choice of mode (i.e., filter function), a simple homodyne measurement with a linear estimator achieves the same error scaling with $N$ as the QFI of the original spin state $|\psi_\varphi\rangle$ at $t=0$.  Further, in the large-$N$ limit, the optimized homodyne SNR converges to half the SNR of an idealized projective measurement $\hat{J}_\alpha$ measurement of the initial spin state.  This is true for the initial state corresponding to a simple Ramsey measurement (in which case we get SQL scaling $\Delta \parameter \sim 1 / \sqrt{N}$, and for the case where the initial state was produced by an optimal twist-untwist spin-squeezing protocol (in which case we get HL scaling, $\Delta \parameter \sim 1 / N$). 


We can state these results more precisely by numerically evaluating Eq.~(\ref{eq:factorization_snr}) in the limit $N \gg 1$.  For the case of an initial state produced by a Ramsey sequence (i.e., a parameter generator $\generator = \hat{J}_y$, c.f.~Eq.~(\ref{eq:initial_state_separable})), we find:
\begin{equation}
\begin{split}
    \text{SNR}[w_{\text{SNR}}(t)] &\sim \frac{N}{2} = \frac{1}{2} \mathcal{F}_\theta(\ket{\psi^S_\theta}) \\
    &= \frac{1}{2} \text{SNR} [ \hat{J}_x ] \quad \text{for } N \to \infty \; .
\end{split}
\label{eq:limit_SNR_separable}
\end{equation}
In contrast, for an initial state produced from an optimal twist-untwist sequence
(c.f.~Eq.~(\ref{eq:initial_state_entangled})), we obtain:
\begin{equation}
\begin{split}
    \text{SNR}[w_{\text{SNR}}(t)] &\sim \frac{1}{2e} N^2 = 0.425 \times \mathcal{F}_\theta(\ket{\psi^E_\theta}) \\
    &= \frac{1}{2} \text{SNR} [ \hat{J}_y ] \quad \text{for } N \to \infty \; .
\end{split}
\label{eq:limit_SNR_entangled}
\end{equation}
where $\text{SNR} [ \hat{J}_\alpha ]$ denotes the SNR of an ideal projective measurement of the initial entangled spin state, see App.~\ref{sec:ideal_linear_measurement}. We show the asymptotic convergence of the SNR to these values for both cases in Fig.~\ref{fig:three_modes_comparison}. Note that if one instead performed a homodyne measurement using the maximum-photon mode $w_\text{pop}(t)$ (c.f.~Eq.~(\ref{eq:most_populated_mode})), one obtains the same scaling behaviour with $N$ with a slightly worse prefactor (see Fig.~\ref{fig:three_modes_comparison} again). The same is true if one uses the maximum-signal mode  $w_\text{sig}(t)$ (c.f.~Eq.~(\ref{eq:most_sensitive})). However, these conclusions change dramatically if one now considers the case where the cavity is detuned from the spin ensemble (something that might be necessary to slow down the superradiant decay to a reasonable timescale). As we discuss in the next section, the differences in the SNR for the maximum-SNR mode and the maximum-photons become much larger in this case.



\begin{table}[t!]
\centering
\setlength{\tabcolsep}{6pt} 
\begin{tabular}{@{}l|ll@{}}
\toprule
\textbf{SNR} & \textbf{$\hat{J}_\alpha$ meas.} & \textbf{\makecell[l]{Optimized \\ homodyne}} \\
\midrule
\makecell[l]{\textbf{Standard} \\ \textbf{Ramsey}, Eq.~\eqref{eq:initial_state_separable}} & $N$, Eq.~\eqref{eq:SNR_projective_separable} & $\frac{N}{2}$, Eq.~\eqref{eq:limit_SNR_separable} \\
\midrule
\textbf{Twist-untwist}, Eq.~\eqref{eq:initial_state_entangled} & $\frac{N^2}{e}$, Eq.~\eqref{eq:SNR_projective_entangled} & $\frac{N^2}{2e}$, Eq.~\eqref{eq:limit_SNR_entangled} \\
\bottomrule
\end{tabular}
\caption{\justifying Summary of the signal-to-noise ratio of the ideal measurements on the spin ensemble (non achievable experimentally because of the inefficient readout) vs. the SNR of the integrated homodyne current measured on the optimal temporal mode $w_{\text{SNR}}(t)$ in Eq.~\eqref{eq:snr_optimizer}. These results, which hold asymptotically in $N$, are already approached for $N\sim\mathcal{O}(10^2)$.}
\label{tab:sensing_strategies}
\end{table}

\subsection{Effective information concentration: optimal-information mode throws away most photons}

We now discuss another remarkable feature of our optimized homodyne protocol:  the optimized homodyne measurement involves a temporal mode whose average photon number scales {\it sublinearly} with the number of spins $N$.  The implication is that for large $N$, even though we measure a tiny fraction of the emitted photons, we retrieve almost all of the information on $\parameter$ in the initial state.  An equivalent statement is that almost all of the output field can be discarded without degrading the precision of the estimation of $\varphi$.  

We can make the above statement more precise.  
Using Eq.~\eqref{eq:photon_number}, we compute the average photon number in the maximum-SNR temporal mode given in Eq.~\eqref{eq:snr_optimizer}.  We use this to define the {\it fractional occupation} of the maximum-SNR mode, $\bar{p}_{\rm opt}$, which varies from $0$ to $1$:
\begin{equation}
    \bar{p}_{\rm opt} \equiv \frac{\bar{n}_{w_{\rm SNR}}}{N} = \frac{1}{N} \int_{0}^\infty \dd t' \dd t \, w_{\rm SNR} (t) g_1 (t, t') w_{\rm SNR}(t') \; . \label{eq:ratio_num_photons}
\end{equation}

As shown in Fig.~\ref{fig:SNR_comparison} (right panel), we find that the fractional occupation of the maximum-SNR mode decays as a power law with increasing $N$; numerical fits suggest a power law decay $\bar{p}_{\rm opt} \propto N^{-0.37}$, implying that the maximum SNR mode has an average photon number $\bar{n}_{\rm opt} \sim N^{0.63}$.  We stress that this is also a ``universal" result, as this scaling does not depend on the choice of the parameter generator $\generator$. 

We note that the small photon number (and hence power) associated with the optimal information mode could be advantageous in some experimental platforms, where one is perhaps worried about possible detector saturation effects associated with the large total photon number and power of the full superradiant burst.  It is also worth stressing that there are many classical analogues of the behaviour we find here, i.e., situations with structured Gaussian noise where the optimal matched filter yields a form that throws away most of the power in the detected field in order to exploit the presence of correlated noise. 
See App.~\ref{sec:interpretation_mode} for more details about the interpretation of the optimal temporal mode of the radiation $w_\text{SNR}(t)$.

\subsection{Practical considerations}

Here we briefly discuss the practical considerations relevant to potential experimental realizations of our general scheme. We first clarify what sets the phase reference for our optimal homodyne measurement.  
For standard homodyne protocols, one works with a coherently driven system, and hence there are natural amplitude and phase quadratures that are defined relative to the phase of the average input field amplitude $\langle \hat{b}_{\rm in}\rangle $.  In such situations, the local oscillator reference is typically derived from the same laser (or source) used to generate the coherent drive $\hat{b}_{\rm in}$. 
In contrast, in our setup there is no coherent photonic driving, and hence no naturally-defined amplitude and phase quadrature.  Instead, the phase reference for the local oscillator will be determined by the same laser (or microwave source) used to encode information in the initial spin state.  This encoding step is phase-sensitive, and the same drives that determine this encoding phase could be used to set the local-oscillator phase.  Formally, the choice of the encoding phase is expressed in Eq.~\eqref{eq:universal_expression_signal}:  the overall phase of the signal is determined by the phase of the matrix element $\singlebra{m_0}\generator\singleket{m_0-1}$ of the generator $\hat{G}$.

Next,  we would like to show that our scheme is robust against detection efficiency $\eta$. 
As is standard, we model a finite efficiency via an imbalanced beam-splitter, and the field entering homodyne detection by $\hat{b}_{\rm det} = \sqrt{\eta}\hat{b}_{\rm out}+\sqrt{1-\eta}\hat{b}_{\rm vac}$, with $\hat{b}_{\rm vac}$ the vacuum noise entering the second input port of the beam-splitter.
Using this expression in the power SNR formula given in Eq.~(\ref{eq:definition_SNR_M}), we obtain
\begin{equation}
     \operatorname{SNR}_{\eta}[u(t)]=
        \frac{ \eta| \bar{S}_u |^2}{\eta\mathrm{Var}(\hat{M}_{u}) + (1-\eta)/2}.
\end{equation}
In our case, the maximum SNR mode has an average photon number $\bar{n}_{\rm opt} \sim N^{\frac{2}{3}}$, and has no squeezing correlations.  As such, the fluctuations of the max-SNR mode $\mathrm{Var}(\hat{M}_{u})$ also scale as $\sim N^{\frac{2}{3}}$, and for $N \gg 1$ will completely dominate over the vacuum noise term in the denominator.  This leads to $\operatorname{SNR}_{\eta}\approx \operatorname{SNR}_{\eta=1}$ if $\eta$ is not too small.

Finally, we note that for large $N$, one does not need phase sensitive detection schemes: a simple heterodyne scheme is also almost ideal (easily implemented in microwave systems using a standard $IQ$ mixer).  This would correspond to the above imperfect homodyne SNR expression with an efficiency $\eta=0.5$.
For large $N$, the same reasoning as above holds: the fluctuations of the max-SNR mode completely dominate the vacuum noise contribution associated with $\eta < 1$, and hence the heterodyne SNR is nearly the same as homodyne SNR.


\section{Including collective spin-spin interactions: case of a detuning spin-ensemble cavity setup}
\label{sec:cavity_detuning}
We now extend our analysis to cover collective dynamics that has the full form described by Eq.~(\ref{eq:lindbladian_decay}), i.e., both the collective superradiant decay dynamics as well as a cavity-mediated spin-spin interaction $\chi$ that has the form of a one-axis twisting Hamiltonian.  This detuned regime is not just a theoretical curiousity: in many systems with large $N$, the resonant case would produce a collective decay rate $\Gamma$ that is too fast to be experimentally accessed.  In such cases, a small detuning $\Delta$ could be used to slow down the superradiant decay to be more compatible with accessible timescales.  While a detuning will reduce the value of $\Gamma$, it also modifies the dynamics as there is now a Hamiltonian spin-spin interation.  We study here whether this modification degrades our basic homodyne measurement strategy.  As we show, the answer is no: and optimized homodyne measurement (using a simple linear estimator) still achieves excellent performance.

\begin{figure*}[tbp]
  \centering
  \includegraphics[width=1.0\textwidth]{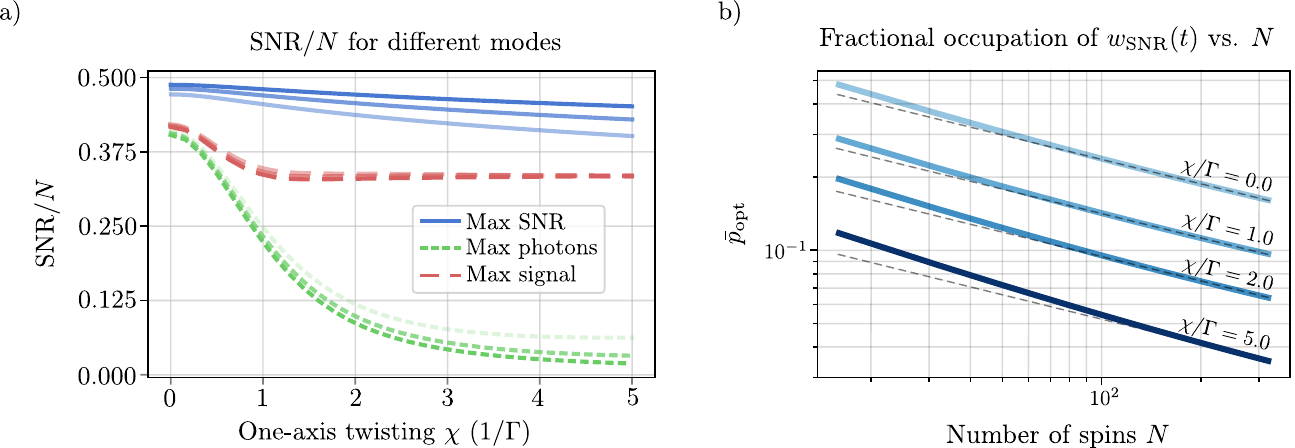} 
  \caption{\justifying a) Homodyne SNR scaled by $N$ for the three temporal modes $w_{\text{SNR}}(t)$ (solid line), $w_{\text{\text{pop}}}(t)$ (short dashed line), $w_{\text{sig}}(t)$ (long dashed line) as a function of the spin-spin interaction $\chi$ produced by a detuned cavity; we consider the unentangled Ramsey-protocol initial state in Eq.~(\ref{eq:initial_state_separable}). For large values of $\chi$, there is a large difference between the SNR of the maximum-SNR mode and the naive maximum-photons mode. The opacity of the lines corresponds to different values of $N$ ($N=32, 64, 128$ for increasing opacity). Note that apart from an overall common scale factor, these curves are the same for any generator $\generator$, and hence the figure also applies to the twist-untwist protocol which achieves Heisenberg scaling. b) We plot the fractional occupation number in Eq.~\eqref{eq:ratio_num_photons}, for the maximum SNR mode in Eq.~\eqref{eq:snr_optimizer}, for an ensemble of spins coupled to a detuned cavity. We fit the power law $r=a N^{b}$ on the results of the numerical computations, finding a consistent result for the exponent $b \simeq -0.37$ as the detuning is varied, and therefore the OAT term $\chi$ changes. The detuning through the OAT term has the effect of making the compression of the information in the few photons of the maximum-SNR mode even more evident.}
\label{fig:SNR_comparison}
\end{figure*}

We repeat the generic mode analysis of the previous sections to dynamics that includes a non-zero interaction $\chi$. One finds changes already at the level of how photons are distributed in the output field.  As $\chi / \Gamma$ becomes appreciable, the photon population becomes less concentrated on just a single mode.  We can again use Eqs.~\eqref{eq:most_populated_mode}, \eqref{eq:most_sensitive}, and \eqref{eq:snr_optimizer} to compute the modes that optimize the average photon number, the net integrated signal, and, most interestingly, the homodyne SNR.  The results for how the SNR of these modes behave are shown in Fig.~\ref{fig:SNR_comparison}, for an initial unentangled spin state corresponding to a Ramsey protocol (c.f.~Eq.~(\ref{eq:initial_state_separable})).  
While the SNR of the maximum-signal and maximum-SNR mode are comparable, the SNR of the most populated mode rapidly drops when the detuning is increased. Because of the factorization in Eq.~\eqref{eq:factorization_snr}, the same functional dependence of the SNR in $\chi$ reported in Fig.~\ref{fig:SNR_comparison} would be valid for every generator $\generator$, when appropriately rescaled with $\bigl| \braket{m_0 \lvert \generator \rvert m_0-1} \bigr|^2$.
We can also study the ``concentration" effect we found for the resonant-cavity case, where the maximum-SNR mode contains a vanishing fraction of the total emitted photons.  We compute numerically the ratio in Eq.~\eqref{eq:ratio_num_photons} for multiple values of the OAT interaction strength $\chi$, each time fitting the decay of the fractional occupation $\bar{p}_{\rm opt}$ with $N$ to a power law.  We observe that the decay exponent varies weakly with the value of $\chi$, with a more steep decay with increasing $\chi$.  This implies that the ``concentration" effect is even more pronounced in the case where there are non-zero spin-spin interactions.  

While the optimized modes $w_{\text{pop}} (t)$, $w_{\text{sig}} (t)$, and $w_{\text{SNR}} (t)$ are all real for $\chi=0$, they become complex function when detuning is tuned up, i.e., $\chi \neq 0$. This implies that the quadrature to measure in the optimal homodyne detection changes non-trivially with time, on the timescale of the (slowed) superradiant emission, making the experimental realization more challenging.

\section{Beyond linear estimation: recovering the full Quantum Fisher information using homodyne measurement}
\label{sec:saturation_qfi}
%

Our analysis so far of homodyne measurement has focused exclusively on the simplest experimental protocol, where the estimate of the parameter $\parameter$ is proportional to the measured, filtered homodyne current associated with a given temporal mode.  As discussed, the shape of the maximum-SNR mode we found in Eq.~\eqref{eq:snr_optimizer}, as well as the corresponding maximum value of the SNR are given by expressions that match the theory of classical generalized matched filters. Once can easily show that {\textit{if}} the homodyne current could be accurately modeled as a Gaussian stochastic process (i.e., its fluctuations involve no higher-order cumulants) than the linear estimator we use is optimal, and the SNR we calculate would be the full Fisher information associated with the homodyne current $I(t)$.  

This then raises an interesting question: is there an appreciable non-Gaussian character to the fluctuations of the homodyne current?  If there are higher cumulants, can these be used to improve the parameter estimation beyond what is possible with a simple linear estimator?  We show that, surprisingly, the answer to both these questions is positive.  
In App.~\ref{sec:optimality_unravelling}, we prove that for the resonant-cavity setup ($\chi = 0$), the full classical Fisher information associated with the full (unfiltered) homodyne current exactly saturates the QFI of the initial spin-ensemble state.  We show this by exploiting a formal equivalence between our setup (where the parameter is encoded in the initial state of the spins), and a continuous-metrology protocol~\cite{yangQuantumCramerRaoPrecision2026}, where the parameter $\parameter$ controls the form of the Lindbladian that governs dynamics. This rigorous equivalence allows us to employ a modified version of the conditions recently formulated in~\cite{mattes_designing_2025} to show
that the full information content of a homodyne measurement (its CFI) saturates the bound set by the QFI of the output field. 

Comparing this result to our results for the SNR, we can conclude that even in the thermodynamic limit of $N \to \infty$, the homodyne current is not a Gaussian process.
In fact we can define a Gaussian stochastic process with average $\mu_\parameter(t) \equiv \langle \hat{I}(t) \rangle_\parameter$, having covariance matrix given by Eq.~\eqref{eq:original_S}. We restrict to the resonant scenario, so that both the signal $\alpha_I(t)$ and the fluctuation matrix are real. To define the Gaussian process we fix a time independent quadrature for the homodyne detection, i.e. $\theta=$const. in Eq~\eqref{eq:homodyne_current}. Also in In App.~\ref{sec:optimality_unravelling} we consider a time independent quadrature for the homodyne detection. In this scenario the CFI of the Gaussian process is exactly the SNR of the linear measurement reported in Eq.~\eqref{eq:optimal_snr_expression}. In summary, the optimal estimator that considers the fluctuations of the current as Gaussian achieves half of the QFI, but we know that considering the full spectrum of correlation in the trajectories would give the full QFI. This means that there are non-Gaussian correlations in the current.

While this result on the full CFI of the homodyne current is both interesting and surprising, making use of it experimentally would require potentially complex processing of the full time-dependent homodyne current. We stress that that the much simpler protocol studied in the main text (based on a simple linear estimator) already comes within a factor of two of the full output-field QFI.  Finally, we note that the same mapping discussed above lets us show that if one does photodetection and uses all the information in the full photo-current $n_{\rm out}(t)$, one can also saturate the QFI of the output field.

\section{Conclusions}
\label{sec:conclusions}

We can summarize the results of work with two key take-home points:
\begin{enumerate}
    \item Filtered homodyne detection with a simple linear estimator is almost optimal for extracting information from a superradiant burst generated by a spin ensemble whose initial state encodes a parameter of interest.   The signal-to-noise ratio of such estimator is only a factor $\simeq 2$ away from the fundamental bound set by the quantum Fisher information.  This result holds for a surprisingly broad class of parameter generators; in particular, it holds for cases where there is Heisenberg scaling in the number of spins $N$.  
    \item A vanishingly small fraction of all the photons in the photonic output field generated by superradiance carries almost all the information encoded in the initial spin state (implying an effective kind of compression of information).  
\end{enumerate}

While our work emphasizes the fundamental aspects of our general setup, we stress that our results could enable a new method for using cavities to readout information stored in a spin ensemble.  Compared to a conventional dispersive readout, the superradiant readout we describe does not require any detuning and hence makes use of the full spin-photon coupling.  This could be advantageous in sensing platforms where solid state spin ensembles couple to microwave cavities.  Here, coupling strengths are weak hence a resonant approach could be highly beneficial.  


Our work also suggests many additional intriguing theory investigations.
For example, could control operations be applied to the spins to influence and perhaps enhance the information of the output state?  Could similar ideas to what we describe be employed in superradiant decay phenomena involving ordered arrays of emitter spins, in the absence of a cavity?  It would also be interesting to develop a more complete understanding of our surprisng finding that the maximum SNR mode contains a vanishing fraction of the emitted photons.  Can one identify an underlying mechanism, and perhaps generalize it to a broader class of systems where this effective compression might be useful?

\section{Code availability}
All the code used to perform the simulations will be made available in a repository upon publication of the manuscript.

\section{Acknowledgments}
This material is based upon work supported by the U.S. Department of Energy Office of Science National Quantum Information Science Research Centers as part of the Q-NEXT center.  Q-NEXT provided primary funding support for F.B. during the completion of this research.  A.C. acknowledges support from the Simons Foundation through a Simons Investigator Award (Grant No. 669487).

\bibliography{reference}

\appendix

\section{Signal-to-noise ratio of ideal linear measurements on the spin ensemble}
\label{sec:ideal_linear_measurement}
In this section we want to set a baseline for the performances that we hope to achieve in measuring the radiation with homodyne detection. We compute in the following the signal-to-noise ratio of linear measurements of the spin ensemble, which we compare to the results of the measurements on the radiation in Table~\ref{tab:sensing_strategies}.

In Eq.~\eqref{eq:initial_state_separable} and Eq.~\eqref{eq:initial_state_entangled} we have characterized the QFI of the two initial states $\ket{\psi_\parameter^S}$ and $\ket{\psi_\parameter^E}$. 
In this appendix, we ask whether this QFI can be achieved via a simple direct measurement on the spin ensemble. We restrict this analysis to measurements of one component of the ensemble's collective spin $\hat{J}_\alpha$, and to the simplest kind of linear estimator, where the value of $\parameter$ is directly estimated from the average of $\hat{J}_\alpha$. This will help to set our expectations correctly when we turn to the superradiance-based method of extracting information.  

For the unentangled state $\ket{\psi_\parameter^S}$, it is obvious that such a simple projective measurement and linear estimator is optimal. This can be seen explicitly by computing the signal-to-noise ratio associated with a projective $\hat{J}_x$ measurement:
\begin{equation}
    \text{SNR}(\ket{\psi^{S}_\parameter}; \hat{J_x}) = \frac{\big| \braket{m_0 | [\hat{J}_y, \hat{J}_x ] | m_0}\big|^2}{\braket{m_0| \hat{J_x}^2 |m_0}} = N \; .
    \label{eq:SNR_projective_separable}
\end{equation}
The SNR coincides with the full QFI of the state, indicating that projective measurements with a linear estimator are ideal.  

We contrast this result with the case of the twist-untwist state $\ket{\psi_\parameter^E}$.  
In this case, the optimal spin component to measure is $\hat{J}_y$, and we find that the SNR for a projective measurement is
\begin{equation}
    \text{SNR}(\ket{\psi^E_\parameter}; \hat{J_y}) = \frac{\big| \braket{m_0| [\hat{U}_s \hat{J}_y \hat{U}_s^\dagger, \hat{J}_y ] | m_0}\big|^2}{\braket{m_0| \hat{J_y}^2 |m_0}}\; .
\end{equation}
Numerically, we observe that
\begin{equation}
    \text{SNR}(\ket{\psi^E_\parameter}; \hat{J}_y) \sim \frac{N^2}{e} = 0.85 \, \mathcal{F}(\ket{\psi_\theta^E}) \quad \text{as } N \to \infty \; .
    \label{eq:SNR_projective_entangled}
\end{equation}
with the asymptotic regime being achieved already for $N\sim \mathcal{O}(10^2)$. In Ref.~\onlinecite{davis_approaching_2016}, it has been proven that this is indeed the maximum SNR for the twist-untwist strategy with a linear estimator. Hence, we see that, in this case, a simple projective measurement (and a linear estimator) allows one to achieve Heisenberg scaling, but it does not saturate the QFI. If we were to consider the full statistics of the outcome of the measurement on $\hat{J}_y$, the we could saturate the QFI. The optimal measurement can be computed from the theory of quantum metrology as the projector onto the eigenvectors of the symmetric logarithmic derivative, see~\cite{toth_quantum_2014}.

\section{Computation of the homodyne signal for the spin ensemble.}
\label{sec:analytical_signal}
Let us start with the definition of $\alpha_I(t)$ in Eq.~\eqref{eq:homodyne_sensitivity},
\begin{equation}
\begin{aligned}
    \alpha_I(t) &= - \sqrt{\Gamma} \braket{m_0| [ \generator, \hat{J}_{-} (t) ] | m_0} \; .
\end{aligned}
\end{equation}
We reformulate the expectation value of the commutator:
\begin{align}
    \braket{[\generator, \hat{J}_{-} (t)]} &= \tr{\hat{\rho} [\generator, \hat{J}_{-}(t)]} \\
    &= \tr{\hat{\rho} \generator \hat{J}_{-} (t) - \generator \hat{\rho} \hat{J}_{-} (t)} \\
    &= \tr{[\hat{\rho}, \generator] \hat{J}_{-} (t)} \\
    &= \tr{\myexp{\mathcal{D} [\sqrt{\Gamma} \hat{J}_{-}] t} \left( [\hat{\rho}, \generator] \right) \hat{J}_{-}} \; .
\end{align}
We assumed that the generator $\generator$ of the signal is collective, i.e., it is entirely defined as an operator by its matrix elements in the symmetric subspace:
\begin{equation}
    \generator = \sum_{m', m = -m_0}^{m_0} \braket{m'|\generator|m} \ket{m'} \! \bra{m} \equiv \sum_{m', m} G_{m', m} \ket{m'} \! \bra{m} \; ,
\end{equation}
with $m_0=\frac{N}{2}$. The commutator $[\hat{\rho}, \hat{G}]$ can therefore be expressed as
\begin{multline}
    [\hat{\rho}, \generator] = \singleket{m_0} \singlebra{m_0} \left( \sum_{m', m = -m_0}^{m_0} G_{m', m} \ket{m'} \! \bra{m} \right) \\ -
    \left( \sum_{m', m = -m_0}^{m_0} G_{m', m} \ket{m'} \! \bra{m} \right) \singleket{m_0} \singlebra{m_0} \; ,
\end{multline}
which yields
\begin{equation}
\begin{split}
    \hat{X} (t=0) \equiv [\hat{\rho}, \generator] &= \sum_{m=-m_0}^{m_0} G_{m_0, m} \ket{m_0} \bra{m} \\
    &\quad - \sum_{m=-m_0}^{m_0} G_{m, m_0} \ket{m} \bra{m_0} \; .
\end{split}
\label{eq:initial_X}
\end{equation}

This operator admits the matrix representation
\begin{widetext}
 \begin{equation}
    \hat{X} (t=0) \equiv \begin{bmatrix}
    0 & 0 & 0 & \cdots & -G_{-m_0, m_0} \\
    0 & 0 & 0 & \ddots & \vdots \\
    0 & 0 & 0 & \ddots & -G_{m_0-2, m_0} \\
    \vdots & \ddots & \ddots & \ddots & -G_{m_0-1, m_0} \\
    G_{m_0, -m_0} & \cdots & G_{m_0, m_0-2} & G_{m_0, m_0-1} & 0
    \end{bmatrix} \; .
    \label{eq:zero_time_commutator}
\end{equation}   
\end{widetext}
We define the time-evolved commutator
\begin{equation}
     \hat{X}(t) = \myexp{\mathcal{D} [\sqrt{\Gamma} \hat{J}_{-}] t} \left( [\hat{\rho}, \generator] \right) = \begin{bmatrix}
    0 & \square & \square & \cdots & \square \\
    \blacksquare & 0 & \Box & \ddots & \vdots \\
    \square & \blacksquare & 0 & \ddots & \square \\
    \vdots & \ddots & \ddots & \ddots & \Box \\
    \square & \cdots & \square & \blacksquare & 0
    \end{bmatrix} \; .
    \label{eq:evolved_commutator}
\end{equation}
This evolution does not introduce any extra coherence in the system, which also implies that the operator $\hat{X}(t)$ has zeros on the diagonal. 
Note that the presence of a Hamiltonian term $H=\chi J_z^2$, which may originate from detuning the spins from the cavity, has the effect of adding a time-dependent phase to all matrix elements. 
However, this does not change the structure of Eq.~\eqref{eq:evolved_commutator} and it will not change our conclusions in the section, so that it can be neglected in the following. The multiplication with
\begin{equation}
    \hat{J}_{-} = \sum_{m = -m_0}^{m_0} \sqrt{m_0(m_0+1) - m(m-1)} \ket{m-1} \! \bra{m} \; .
\end{equation}
selects only the lower diagonal of the time-evolved commutator as terms that contribute to the homodyne signal $\alpha_I(t)$, which have been indicated with the black boxes in Eq.~\eqref{eq:evolved_commutator}. 
Therefore, we only need to track the evolution of the $N$ elements of the subdiagonal of the commutator, which we indicate with $X_{m} (t)$ for $m=-j, -j+1, \cdots, j-1$. These elements satisfy the differential equation
\begin{equation}
    \frac{\dd X_m(t)}{\dd t} = -\frac{1}{2} \left( \Gamma_m + \Gamma_{m+1} \right) + \sqrt{\Gamma_m \Gamma_{m+1}} X_{m+1} \; ,
    \label{eq:equation_motion}
\end{equation}
with $\Gamma_m \equiv \Gamma (m+m_0) (m_0 - m + 1)$, and $X_{m_0} (t) \equiv 0\,, \; \forall t$ by definition. The initial state is
\begin{equation}
    X_m(t=0) = \delta_{m, m_0-1} G_{m_0, m_0 - 1} \; ,
    \label{eq:initial_value_X_operator}
\end{equation}
The final expression for the homodyne signal is
\begin{equation}
    \alpha_I(t) = \sum_{m=-m_0}^{m_0-1} X_m (t) \sqrt{(m+m_0+1) (m_0-m)} \; .
    \label{eq:formula_alpha}
\end{equation}

\section{Proof of the separation theorem}
\label{app:separation_theorem}
\subsection{Separation theorem without cavity}
We will now prove that the homodyne signal for an ensemble of spins starting in the fully excited state is factorized, as written in Eq.~\eqref{eq:universal_expression_signal}. This statement follows from Eq.~\eqref{eq:formula_alpha} in App.~\ref{sec:analytical_signal}: The expression for the signal involves $X_m(t)$, which evolves linearly in time from an initial condition proportional to a single entry of the signal generator $G$, see Eq.~\eqref{eq:initial_value_X_operator} and Eq.~\eqref{eq:equation_motion}; we can indeed write 
\begin{equation}
    X_m(t=0) = G_{m_0, m_0-1} \tilde{X}_m (t=0) \; ,
\end{equation}
with $\tilde{X}_m (t=0) = \delta_{m, m_0-1}$ and $G_{m_0, m_0-1} \equiv \singlebra{m_0}\generator\singleket{m_0-1}$. The renormalized vector $\tilde{X}(t)$ evolves according to the same equation of motion reported in Eq.~\eqref{eq:equation_motion} for $X(t)$. By linearity we can write the expression for the homodyne signal as a function of  $\tilde{X}_m(t)$:
\begin{equation}
    \alpha_I(t) = G_{m_0, m_0 - 1} \sum_{m=-m_0}^{m_0-1} \tilde{X}_m (t) \sqrt{(m+m_0+1) (m_0-m)} \; ,
\end{equation}
which proves the theorem with
\begin{equation}
    \tilde{\alpha}_I(t) \equiv \sum_{m=-m_0}^{m_0-1} \tilde{X}_m (t) \sqrt{(m+m_0+1) (m_0-m)} \; .
\end{equation}

\subsection{Separation theorem with cavity}
In this paragraph, we prove the same separation statement on $\alpha_I(t)$ for a system with a cavity that has not been adiabatically eliminated. We are able to prove the separation theorem for whatever collective Hamiltonian acting on the spins. We assume the interaction term between the cavity and the spins to be that of the Tavis-Cummings model in Eq.~\eqref{eq:hamiltonian_tavis_cummings}, while the cavity loss is described by the superoperator
\begin{equation}
    \mathcal{D} [\sqrt{\kappa} \hat{a}](\hat{X}) \equiv \kappa \left( \hat{a} \hat{X} \hat{a}^\dagger - \frac{1}{2} \lbrace \hat{a}^\dagger \hat{a} , \hat{X} \rbrace \right) \; .
    \label{eq:cavity_loss}
\end{equation}
We will use the shortened notation $\ket{m} \equiv \singleket{j=N/2, m}$, because the spin ensemble remains in the symmetric subspace during the evolution, since the interaction with the cavity is collective and we assume a collective Hamiltonian. We indicate with $\ket{n}$ the Fock state of the cavity, and with $\ket{m, n}$ the joint state of the spins an the cavity containing $n$ photons.  We start by analyzing the evolution of an operator $\hat{X}(t=0) \equiv \doubleket{M}{N} \! \doublebra{M'}{N'}$, under the Tavis-Cummings model. If the total energy of the spins and the cavity is conserved, in the evolved operator $\hat{X}(t)$, which we write as
\begin{equation}
    \hat{X}(t) =\sum_{\substack{n, m \\ n', m'}} 
    X_{n'm', nm}(t) \doubleket{m}{n} \! \doublebra{m'}{n'} \; ,
\end{equation}
only those terms satisfying
\begin{equation}
    \begin{aligned}
    m + n   &= M + N \, , \\
    m' + n' &= M' + N' \, , 
    \end{aligned}
\end{equation}
can have a non-zero coefficient. We now consider also the loss term and we evolve $\hat{X}(t)$ under the master equation for the decaying ensemble of spins. The commutator with the photon number in Eq.~\eqref{eq:cavity_loss} acts on each operator $\doubleket{m}{n} \! \doublebra{m'}{n'}$ by sending it to $ (n+n') \doubleket{m}{n} \! \doublebra{m'}{n'}$, thereby it doesn't create new non-zero terms in the evolution of $\hat{X}(t)$, and in particular it is energy preserving. On the other hand $\hat{a} \hat{X} \hat{a}^\dagger$ is not energy preserving and it destroys quanta of energy on both the kets and bras of $\doubleket{m}{n} \! \doublebra{m'}{n'}$. The application of the Hamiltonian term again exchanges energy between the cavity and the spins, so that as a whole the energy of the system (ensemble and cavity) can only decrease, but separately for the kets and the bras of the generic operator $\hat{X}(t)$. We conclude that in evolving $\hat{X}(t)$ we can have only non-zero coefficients $X_{n'm', nm}\neq0$ if the indices satisfy 
\begin{equation}
    \begin{aligned}
        m + n   &\le M + N \; ,\\
        m' + n' &\le M' + N' \; ,
    \end{aligned}
\end{equation}
The final key observation is, that, in the photon loss process described by Eq.~\eqref{eq:cavity_loss}, two quanta of energy are annihilated at the same time on the left and on the right of $\doubleket{m}{n} \! \doublebra{m'}{n'}$, which implies that the difference in the number of excitations (between cavity and spins) on the bras and the kets is to remain constant. We have therefore the additional constraint
\begin{equation}
    m + n - (n' + m') = M + N - (N' + M') \; .
\end{equation}
We now introduce an expression for the homodyne signal $\alpha_I(t)$, analogous to Eq.~\eqref{eq:homodyne_sensitivity}, but without the adiabatic elimination of the cavity. Starting from the input output relation
\begin{equation}
    \bout(t) = \sqrt{\kappa} \hat{a}(t) + \bin (t) \; ,
\end{equation}
we can write
\begin{equation}
    \alpha_I (t) = \sqrt{k} \langle [\generator \otimes \id , \hat{a}(t) ] \rangle \; .
\end{equation}
The generator of the encoding $\generator$ acts only on the spins, while $\hat{a}(t)$ is the evolved decay operator acting both on the spins and the cavity mode. We focus on the computation of $\langle [\generator \otimes \id , \hat{a}(t) ] \rangle$, from which the full homodyne signal $\alpha_I(t)$ can be obtained. This term can be rewritten as
\begin{equation}
    \langle [\generator \otimes \id , \hat{a}(t) ] \rangle = \tr{\hat{X}(t) \, \hat{a}} \; ,
\end{equation}
where the operator $\hat{X}(t)$ is the evolution of the commutator $[\hat{\rho} \otimes \ket{0} \! \bra{0}, \generator \otimes \id]$, with the Lindbladian evolution $e^{\mathcal{L} t}$ for the dissipative Tavis-Cummings model, constructed with the Hamiltonian in Eq.~\eqref{eq:hamiltonian_tavis_cummings} and the dissipator in Eq.~\eqref{eq:cavity_loss}, i.e
\begin{equation}
    \hat{X}(t) = \myexp{\mathcal{L} t} \left( [\hat{\rho} \otimes \ket{0} \! \bra{0}, \generator \otimes \id] \right) \; .
\end{equation}
We work under the assumption that the initial state of the spins is $\hat{\rho} \equiv \ket{m_0} \! \bra{m_0}$, and that the cavity is initially in the vacuum state, so that we can specify the commutator:
\begin{multline}
     [\hat{\rho} \otimes \ket{0} \! \bra{0}, \generator \otimes \id] = 
     \underbrace{\sum_{M'=-m_0}^{m_0} \doubleket{m_0}{0} \! \doublebra{M'}{0} G_{M', m_0}}_{\text{I}}
        \; \\ - \;
     \underbrace{\sum_{M=-m_0}^{m_0} G_{m_0, M} \doubleket{M}{0} \! \doublebra{m_0}{0}}_{\text{II}} \; .
     \label{eq:commutator_cavity}
\end{multline}
We also write the lowering operator of the cavity in the Fock basis, while adding explicitly the identity on the spins:
\begin{equation}
    \id \otimes \hat{a} = \left( \sum_{m=-m_0}^{m_0} \singleket{m} \! \singlebra{m} \right) \otimes \left( \sum_{n=0}^\infty \sqrt{n} \singleket{n-1} \! \singlebra{n} \right) \; .
    \label{eq:decay_operator_system}
\end{equation}
Let us consider the evolved version of the operator in Eq.~\eqref{eq:commutator_cavity}. We will focus only on those matrix elements that produce a non-zero value when contracted with $\id \otimes \hat{a}$. We divide this analysis for the elements of summand I and II in Eq.~\eqref{eq:commutator_cavity}.
\begin{itemize}
    \item I. $m+n - m' - n' = j - M'$. In order for $\doubleket{m}{n} \! \doublebra{m'}{n'}$ to give a non-zero signal in $\tr{ \doubleket{m}{n} \! \doublebra{m}{n} \, a}$ we need $n-n'=1$ and $m=m'$, which fixes $M' = j - 1$, i.e. only a single operator in the first summation is non-zero.
    \item II. $m+n - m' - n' = M - j$. The same analysis gives $M = j + 1$, which is out of the range of the summation, and it means that the contribution of II to the homodyne signal is zero.
\end{itemize}
We conclude that the homodyne signal is:
\begin{equation*}
    \alpha_I(t) = G_{m_0, m_0-1} \trsquare{ \myexp{\mathcal{L} t} \left( \doubleket{m_0}{0} \! \doublebra{m_0-1}{0} \right) \cdot \hat{a} } \; ,
\end{equation*}
which proves the separation theorem.

\section{Definitions of the estimator and its performance}
\label{sec:definition_information_quantities}
In this appendix, we define the relevant local information quantities we use to establish the metrological performance of our homodyne detection protocol. We consider measuring a Hermitian observable $\hat{M} \equiv \int_{\mu \in \mathcal{M}} \dd \mu \, x \ket{\mu} \! \bra{\mu}$ on the state $\hat{\rho}_R^\parameter (t)$ of the output field at time $t$, which carries a dependence on $\parameter$. 
We denote by $\mu \in \mathcal{M}$ the different possible measurement outcomes, which are observed with probability
\begin{equation}
    p(\mu|\parameter) \equiv \tr{\hat{\rho}_R^\parameter (t) \ket{\mu} \! \bra{\mu}} \; ,
\end{equation}
%
We omit the time dependence in the probability distribution $p(\mu|\parameter)$ to make the notation easier. The value of the parameter $\parameter$ that is encoded in the state of the spins by the unitaries in Eq.~\eqref{eq:initial_state_separable} and Eq.~\eqref{eq:initial_state_entangled} is $\parameter = \parameter_0$. 
Since we work in the regime of local estimation, we will study the estimation of small deviations of the parameter $\parameter$ from the reference value $\parameter_0$. 
For clarity, we distinguish between the encoded parameter $\parameter$ and its value $\parameter_0$ in this section. In the rest of the manuscript, however, we drop the subscript in $\parameter_0$ and use the symbol $\parameter$ also for actual value encoded in the spin ensemble. 
While, in this section, we consider observables measured on the state of the radiation field, it is always possible to reexpress them as expectation values measured on the state of the spins by means of the input-output theory, as we do throughout the manuscript and in the appendix (see also~\cite{law_dynamic_2007}). 
The expectation value of an arbitrary Hermitian observable $\hat{A}$ with respect to $\hat{\rho}_R^\parameter(t)$ is $ \langle \hat{A} \rangle_{\parameter} \equiv \tr{\hat{A}\hat{\rho}_R^\parameter (t)}$ (where, again, we omit the time dependence for the sake of clarity).
The average of the observable $\hat{M}$ and its square $\hat{M}^2$ can then be expressed as
\begin{align}
    \langle \hat{M} \rangle_{\parameter} 
    &=  \int_{\mu \in \mathcal{M}} \dd \mu \, \mu \,  p(\mu|\parameter)  \; , \\
    \langle \hat{M}^2 \rangle_{\parameter}
    &= \int_{\mu \in \mathcal{M}} \dd \mu \, \mu^2 \, p(\mu|\parameter) \; .
\end{align}
%
The variance of the observable is defined as
\begin{align}
    \mathrm{Var}(\hat{M})_{\parameter} 
    &\equiv \langle \hat{M}^2 \rangle_\parameter 
       - \bigl\langle \hat{M} \bigr\rangle_{\parameter} ^2 \\
    &= \mathbb{E}_{\parameter} \left[ \mu^2 \right] - 
        \left( \mathbb{E}_{\parameter} \left[ \mu \right] \right)^2 \; ,
\end{align}
where the expectation value over $p(\mu|\parameter)$ is
\begin{equation}
    \mathbb{E}_\parameter [ \cdot] \equiv \int_{\mu \in \chi} \mathrm{d}\mu\cdot  p(\mu|\parameter) \; .
\end{equation}
In order to estimate the parameter $\parameter$, multiple experiments are repeated, measuring $\hat{M}$ $K$ times, and obtaining $K$ random measurement outcomes $\lbrace \mu_k \rbrace_{k=1}^K$. It is possible to construct an estimator for the expectation value $\langle M\rangle_\parameter$ from the empirical average of the observed results of the experiment, i.e.
\begin{equation}
     \bar{x} \equiv \frac{1}{K} \sum_{k=1}^K \mu_k \; .
     \label{eq:empirical_average}
\end{equation}
In the asymptotic limit $K\rightarrow\infty$, it converges in probability to the expectation value of the observable,
\begin{equation}
    \lim_{K\rightarrow\infty} \bar{\mu} = \bigl\langle \hat{M} \bigr\rangle_{\parameter} \; .
\end{equation}
It is possible to construct a locally unbiased estimator~\cite{hayashi_asymptotic_2005} for the parameter of interest at the position $\parameter_0$ from the average $\bar{x}$, i.e.,
\begin{equation}
    \check{\parameter} \equiv \left( \frac{\partial \langle \hat{M} \rangle_\parameter}{\partial \parameter} 
    \Bigg|_{\parameter = \parameter_0} \right)^{-1} \left( \bar{\mu} - \langle \hat{M} \rangle_{\parameter_0} \right) + \parameter_0 \; ,
    \label{eq:estimator_varphi}
\end{equation}
which is sensitive to deviations from the reference value $\parameter_0$. It is easy to verify
\begin{align*}
    \mathbb{E}_{\parameter_0} \left[ \check{\parameter} \right] = \parameter_0 
    \; \text{and} \; 
    \frac{\partial \mathbb{E}_{\parameter} \left[ \check{\parameter} \right]}{\partial \parameter} = 1 \; ,
\end{align*}
which are the conditions for local unbiasedness. The precision of $\check{\parameter}$ is gauged by the signal-to-noise ratio:
\begin{equation}
    \mathrm{Var} \left( \check{\parameter} \right)_{\parameter_0} = \frac{1}{{\mathrm{SNR}}} \; , \quad 
    \mathrm{SNR} \equiv \frac{\left|  \frac{\partial \langle \hat{M} \rangle_\parameter}{\partial \parameter} 
  \right|^2_{\parameter = \parameter_0}}{{\mathrm{Var}(\hat{M})_{\parameter_{0}}}} \; .
    \label{eq:def_snr}
\end{equation}
Upper bounds to the SNR are the classical Fisher information (CFI, computed from the classical probability distribution $p(\mu|\varphi)$) and the quantum Fisher information (QFI) of $\hat{\rho}_R^\parameter (t)$ (obtained by maximizing the CFI over all possible measurements), both evaluated at $\parameter_0$~\cite{pezze_quantum_2018}. 
The QFI will be our reference value for the maximum amount of information that can be extracted from the radiation. 
Note that the joint evolution of the spins, the cavity, and the radiation field is unitary and the encoding happens entirely before the decay. 
Therefore, in the absence of unobserved decay channels, the QFI of the state of the radiation field long after the decay has ended must be equal to the QFI of the spin ensemble right after the encoding, i.e., the QFI of the state given in Eq.~\eqref{eq:generic_initial_state}. 

For a homodyne detection measurement, we have $\hat{M} = \hat{M}_u$ as defined in Eq.~\eqref{eq:observable}.

\section{Derivation of the photon number sensitivity}
\label{sec:photon_number_sensitivity}
In this appendix we calculate the sensitivity of a photon counting measurement applied to a single mode of the radiation emitted by the spin ensemble. The expressions derived in this section are applicable to every generator $\generator$ of the parameter $\parameter$. The temporal-mode formalism allows us to classify the temporal modes of the radiation field according to their sensitivity under a photon-counting measurement. We see how for a superradiant ensemble of spins initialized in the fully excited state, the sensitivity of the photon number is always zero for whatever metrological generator $\generator$. Our analysis mirrors the derivation in~\cite{law_dynamic_2007}. 

\textit{Parametric derivative of the correlation function}. We want to compute the sensitivity of the photon number in a generic mode. We introduce explicitly the dependence of the photon number on $\parameter$, and we write the average number of photons in a temporal mode of the output field:
\begin{equation}
    \bar n_u(\parameter) \equiv \langle \hat{b}^\dagger_u \hat{b}_u \rangle_\parameter =  \int_{0}^\infty \dd t' \dd t \, u^{*} (t') g_1 (t', t ; \parameter) u(t) \; .
\label{eq:photon_number_parameter}
\end{equation}
with $g_1(t', t; \parameter) \equiv \Gamma \langle \hat{J}_{+} (t') \hat{J}_{-}(t) \rangle_\parameter$. The expectation value is $\langle \cdot \rangle_\parameter \equiv \langle m_0 | U_\parameter^\dagger \cdot U_\parameter |m_0 \rangle$. The sensitivity of the photon number to variations of $\parameter$ is gauged by the derivative
\begin{equation}
    \frac{\partial \bar n_u(\parameter)}{\partial \parameter} \Biggr|_{\parameter = 0} = \int_{0}^\infty \dd t' \dd t \, u^{*} (t') \frac{\partial g_1 (t', t; \parameter)}{\partial \parameter} \Biggr|_{\parameter = 0} u(t) \; .
\label{eq:photon_number_parameter_derivative}
\end{equation}
It contains the parametric derivative of the correlation function with respect to $\parameter$,
\begin{equation}
    \frac{\partial g_1 (t', t; \parameter)}{\partial \parameter} \Biggr|_{\parameter = 0} = i \Gamma \langle [ \generator, \hat{J}_{+} (t') \hat{J}_{-} (t) ] \rangle \; ,
\label{eq:commutator_corr_function}
\end{equation}
with the expectation value taken on the unencoded initial state $\ket{m_0}$. This expression has the form of a linear-response function for the operator $\hat{J}_{+} (t') \hat{J}_{-} (t)$ under a small perturbation generated by $\generator$. Let us indicate with $\hat{\rho} = \ket{m_0} \! \bra{m_0}$ the initial, unencoded state of the spins, then we can rewrite the derivative as follows.
\begin{align}
    \frac{\partial g_1 (t', t; \parameter)}{\partial \parameter} \Biggr|_{\parameter = 0} &= i \Gamma \, \tr{ \hat{\rho} [ \hat{G}, \hat{J}_{+} (t') \hat{J}_{-} (t) ] } \\
    &= i \Gamma \, \tr{ [\hat{\rho}, \hat{G}] \hat{J}_{+} (t') \hat{J}_{-} (t) } \; .
    \label{eq:parametric_derivative}
\end{align}
We will use this expression in the following paragraph to prove that the sensitivity of the photon counting measurement is always zero.  This result tells us that, with the exception of pathological cases where both the signal and the noise tend to zero at the same rate (which are not practical from an experimental point of view) the photon counting measurement is never useful in superradiant metrology.

\textit{Sensitivity of the photon counting measurement.} We can compute the expression in Eq.~\eqref{eq:evolved_commutator} by starting from the evolution of the commutator $[\hat{\rho}, \generator]$, which is given by the expression in Eq.~\eqref{eq:evolved_commutator}. Given the time-evolved operator $\hat{X}(t')$ and $\hat{J}_{-}$,
\begin{align}
    \hat{X}(t') &= \sum_{m', m = -m_0}^{m_0} X_{m' m} (t') \singleket{m'} \singlebra{m} \; \\
    \hat{J}_{-} &= \sum_{m = -m_0}^{m_0} \sqrt{m_0(m_0+1) - m(m-1)} \ket{m-1} \! \bra{m} \; ,
\end{align}
we evaluate $\hat{J}_{-} X(t')$,
\begin{widetext}
\begin{equation}
     \hat{J}_{-} \hat{X}(t') = \left( \sum_{m'' = -m_0}^{m_0} \sqrt{m_0(m_0+1) - m''(m''-1)} \ket{m''-1} \! \bra{m''} \right) \cdot \left( \sum_{m', m = -m_0}^{m_0} X_{m' m} (t') \ket{m'} \bra{m} \right) \; ,
\end{equation}
\end{widetext}
which, simplified, leaves us with
\begin{multline}
    \hat{J}_{-} \hat{X}(t') = \sum_{m, m' = -m_0}^{m_0} \sqrt{m_0(m_0+1) - m'(m'-1)} \\ \cdot X_{m'm}(t') \ket{m'-1} \! \bra{m} \; .
\end{multline}
This operator then has to be evolved until time $t$. Because $X_{mm}(t) = 0$ the sub-diagonal with $m'=m$ of $\hat{J}_{-} \hat{X}(t')$ remains zero. We can therefore write the evolved state as
\begin{align}
    \hat{Y} &= \myexp{\mathcal{D} [\sqrt{\Gamma} \hat{J}_{-}] (t - t')} \left( \hat{J}_{-} \hat{X}(t') \right) \\ &= \sum_{m, m' = -m_0}^{j} Y_{m' m} \ket{m'} \! \bra{m} \; ,
\end{align}
with $Y_{m'm} = 0$ for $m'= m - 1$. To complete the computation of the correlation function, we multiply by $\hat{J}_+$, defined by
\begin{equation}
     \hat{J}_{+} = \sum_{m = -m_0}^{m_0} \sqrt{m_0(m_0+1) - m(m+1)} \ket{m} \! \bra{m-1} \; ,
\end{equation}
and we obtain
\begin{widetext}
\begin{equation}
    \hat{J}_{+} \hat{Y} = \left( \sum_{m'' = -m_0}^{m_0} \sqrt{m_0(m_0+1) - m''(m''+1)} \ket{m''} \! \bra{m''-1} \right) \\
    \cdot \sum_{m, m' = -j}^{j} Y_{m' m} \ket{m'} \! \bra{m} \; .
\end{equation}
\end{widetext}
We then take the trace of the above expression and we see that two delta functions appear: $\delta_{m''-1, m'} \delta_{m, m''}$, which give $m' = m-1$. Since we had previously established that $Y_{m'm} = 0$, we conclude that the whole expression is zero. 
We thus found the result
\begin{equation}
    \frac{\partial g_1 (t', t; \parameter)}{\partial \parameter} \Biggr|_{\parameter = 0} = 0 \; ,
\end{equation}
which means that photon counting has zero sensitivity. 
Note that it is sufficient to show that the signal is zero to conclude that the measurement is not useful. 
If one wanted to compute the photon-number fluctuations as well, one would need to evaluate the four-time correlation function of the spin ensemble, too.

In App.~\ref{sec:optimality_unravelling}, we see that a photon-counting measurement saturates the quantum Fisher information if the arrival time of the photons is taken into account.

\section{Fluctuations of the filtered integrated homodyne current}
\label{sec:fluctuations_integrated}

We compute the fluctuations of the integrated current $\hat{M}_u$, which appear in the formula for the SNR in Eq.~\eqref{eq:definition_SNR_M}. We can write
\begin{equation}
\begin{split}
    \text{Var}(\hat{M}_u) &= \langle \hat{M}_u^2 \rangle - \langle \hat{M}_u \rangle^2 \\
    &= \int_0^\infty \dd t' \int_0^\infty \dd t \, S_{II, u} (t', t) \; ,
\end{split}
\label{eq:variance_integral}
\end{equation}
with the correlation function defined in Eq.~\eqref{eq:S_II_mode_u}. We can break down the computation of this quantity by starting from the average of the observable, which is zero:
\begin{equation}
    \langle \hat{M}_u \rangle = \int_{0}^\infty \dd t \, \langle I_{u} (t) \rangle  = 0 \; ,
\label{eq:average_Mu}
\end{equation}
due to the fact that the expectation values of both $\hat{J}_{+} (t)$ and $\hat{J}_{-} (t)$ vanish under weak symmetry. When the state is omitted, the expectation value is assumed to be computed with respect to $\ket{m_0}$. Consequently, we only need to compute $\langle \hat{M}_u^2 \rangle$. We proceed by expressing it as a function of the current operator:
\begin{equation}
\begin{split}
    \langle \hat{M}_u^2 \rangle &= \int_0^\infty \dd t' \int_0^\infty \dd t \, \langle \hat{I}_u(t') \hat{I}_u(t) \rangle \\ 
    &= \int_0^\infty \dd t' \int_0^\infty \dd t \, \langle [u(t') \hat{b}_{\text{out}}(t') + u^*(t') \hat{b}_{\text{out}}^\dagger(t')] \\
    &\quad \times [u(t) \hat{b}_{\text{out}}(t) + u^*(t) \hat{b}_{\text{out}}^\dagger(t)] \rangle \; .
\end{split}
\label{eq:fluctuations_full}
\end{equation}
Due to the rotational symmetry of the initial state, we can neglect the squeezing terms $\langle \hat{b}_{\text{out}}(t') \hat{b}_{\text{out}}(t) \rangle$ and $\langle \hat{b}_{\text{out}}^\dagger(t') \hat{b}_{\text{out}}^\dagger(t) \rangle$. We are then left with the expression:
\begin{equation}
\begin{split}
    \langle \hat{M}_u^2 \rangle &= \int_0^\infty \dd t' \int_0^\infty \dd t \, u(t') u^* (t) \langle \hat{b}_{\text{out}}(t') \hat{b}_{\text{out}}^\dagger(t) \rangle \\
    &\quad + \int_0^\infty \dd t' \int_0^\infty \dd t \, u^*(t') u (t) \langle \hat{b}_{\text{out}}^\dagger(t') \hat{b}_{\text{out}}(t) \rangle \; .
\end{split}
    \label{eq:intermediate_proof}
\end{equation}
We observe that the integration domain over $t'$ and $t$ is symmetric under variable exchange. This implies that only the component of the integrand that is symmetric in time contributes to the total integral. Thus, we can exchange the time variables $t' \leftrightarrow t$ in the second summand of Eq.~\eqref{eq:intermediate_proof}, yielding:
\begin{equation}
\begin{split}
    \langle \hat{M}_u^2 \rangle &= \int_0^\infty \dd t' \int_0^\infty \dd t \, u(t') u^* (t) \langle \hat{b}_{\text{out}}(t') \hat{b}_{\text{out}}^\dagger(t) \rangle \\
    &\quad + \int_0^\infty \dd t' \int_0^\infty \dd t \, u^*(t) u (t') \langle \hat{b}_{\text{out}}^\dagger(t) \hat{b}_{\text{out}}(t') \rangle \; .
\end{split}
    \label{eq:second_to_last_step}
\end{equation}
By applying the commutation relations, we obtain the final formula for the fluctuations:
\begin{equation}
\begin{split}
     \langle \hat{M}_u^2 \rangle &= \int_0^\infty \dd t' \int_0^\infty \dd t \, u^*(t) \big[ \delta (t - t')  \\
     &\quad + 2 \langle \hat{b}_{\text{out}}^\dagger(t) \hat{b}_{\text{out}}(t') \rangle \big] u(t') \\
     &= \int_0^\infty \dd t' \int_0^\infty \dd t \, u^*(t) \fluctuation(t, t') u(t') \\
     &= \int_0^\infty \dd t' \int_0^\infty \dd t \, u^*(t') \fluctuation(t', t) u(t) \; ,
\end{split}
\label{eq:final_fluctuations}
\end{equation}
where in the second-to-last step we have used the definition of the fluctuation function $\fluctuation$, and in the final step we have exchanged the time variables via symmetrization.

There is also a second, more direct approach to the proof, which involves the introduction of the operator $\hat{N}_u \equiv \int_0^\infty \dd t \, \hat{I}^\perp_u (t)$. This is the integral of the current operator $\hat{I}^\perp_u(t) = -i [u(t) \hat{b}_{\text{out}}(t) - u^*(t) \hat{b}_{\text{out}}^\dagger(t)]$, which corresponds to the quadrature orthogonal to $\hat{I}_u$. A direct computation using the commutation relations yields:
\begin{equation}
    \langle \hat{M}_u^2 \rangle + \langle \hat{N}_u^2 \rangle = 2 ( 2 \bar{n}_u + 1) \; ,
\label{eq:sum_quadratures}
\end{equation}
where $\bar{n}_u$ is the average photon number of the mode $u(t)$, as defined in Eq.~\eqref{eq:photon_number}. Due to the weak symmetry, the fluctuations of the two orthogonal quadratures are identical, i.e., $\langle \hat{M}_u^2 \rangle = \langle \hat{N}_u^2 \rangle$. This implies:
\begin{equation}
\begin{split}
    \langle \hat{M}_u^2 \rangle &= \int_{0}^\infty \dd t' \int_{0}^\infty \dd t \, u^{*} (t') \big[ 2 g_1 (t', t) \\
    &\quad + \delta(t' - t) \big] u(t) \; ,
\end{split}
\label{eq:fluctuations_g1}
\end{equation}
which, by virtue of the definition in Eq.~\eqref{eq:g_1_t_t} for the two-point correlation function $g_1(t', t)$, gives the fluctuations as a function of $\fluctuation$, noting that $\fluctuation(t', t) = 2 g_1 (t', t) + \delta(t' - t)$.

\section{Analytical computation of the maximum signal-to-noise ratio mode}
\label{sec:optimization_snr}
In this appendix, we apply the Lagrangian multiplier formalism to optimize the SNR of the homodyne measurement, as it is defined in Eq.~\eqref{eq:optimal_mode}. The goal is to find the expression for the temporal mode $w_\text{SNR}(t)$ reported in Eq.~\eqref{eq:snr_optimizer}. We indicate by $f$ the fluctuations of of the mode $w_\text{SNR}(t)$, computed according to Eq.~\eqref{eq:S_spins}:
\begin{equation}
    f \equiv {\int_{0}^{\infty} \dd t' \int_{0}^{\infty} \dd t \, w_\text{SNR}^{*} (t') \fluctuation (t', t) w_\text{SNR}(t)} \; .
    \label{eq:constraint_fluctuations}
\end{equation}
We can reduce the minimization of $u(t)$ to the space of functions that have a fixed level of fluctuations $f$. We will see that we do not need to know the value of $f$ to solve the optimization problem, we merely need to assume that it exists. Let us write the Lagrangian for the maximization of the signal, given the fluctuations:
\begin{equation}
\begin{aligned}
    \mathcal{L}[u] &\equiv 4 \Biggl| \operatorname{Re} \int_{0}^\infty \dd t \, u(t) \alpha_I(t)  \Biggr|^2 \\
    &\quad - \lambda \left( \int_0^\infty \dd t' \int_0^\infty \dd t \, 
    u^*(t') \fluctuation (t', t) u(t) - f \right) \; .
\end{aligned}
\end{equation}
The mode $w_\text{SNR}(t)$ that optimizes the sensitivity for a fixed SNR is the solution to
\begin{equation}
\begin{aligned}
    \frac{\partial \mathcal{L} [u]}{\partial u^{*}(t)}
    &= \alpha_I^{*} (t) \left( 4 \operatorname{Re} \int_0^\infty \dd t \, u(t) \alpha_I(t) \right) \\
    &\quad - \lambda \int_{0}^T \dd t \, \fluctuation (t, t') u(t') = 0 \; .
\end{aligned}
\end{equation}
The scalar product $\int_0^\infty \dd t \, u(t) \alpha_I(t)$ is a constant, albeit dependent on $u(t)$, so that we can write
\begin{equation}
    u(t) \propto \int_{0}^T \dd t \, \fluctuation^{-1} (t, t') \alpha_I^{*} (t') \; .
    \label{eq:solution_optimization}
\end{equation}
After normalization, Eq.~\eqref{eq:solution_optimization} is the mode that optimized the SNR.

An analytical expression for the maximum-SNR mode $w_\text{SNR}(t)$ can be computed also in the case squeezing correlations are present in the state of the spins (and the emitted radiation), i.e. $\langle \bout(t') \bout(t) \rangle \neq 0$, by using the same Lagrangian multiplier formalism. This expression is never needed throughout the manuscript, but we nevertheless report a derivation of this optimized mode here. The Lagrangian with the fluctuations of the integrated homodyne current $\hat{I}_u$, for a mode $u(t)$, in the presence of squeezing correlation in the output field (originating from spin squeezing in the ensemble), can be written as
\begin{equation}
\begin{aligned}
    \mathcal{L}[u] &\equiv 4 \Biggl| \operatorname{Re} \int_{0}^T \dd t \, u(t) \alpha_I(t)  \Biggr|^2 \\
    & \quad - \lambda \left( \int_{0}^{T} \dd t' \int_{0}^{T} \dd t \, u^{*} (t') \fluctuation(t', t) u(t) + \right. \\
    & \quad\quad \int_{0}^{T} \dd t' \int_{0}^{T} \dd t \, u (t') B (t', t) u(t) + \\
    & \quad\quad \left. \int_{0}^{T} \dd t' \int_{0}^{T} \dd t \, u^{*} (t') B^\dagger (t', t) u^{*}(t)  - f \right) \; .
\end{aligned}
\end{equation}
with $B(t', t) \equiv \langle \hat{J}_{-}(t') \hat{J}_{-}(t) \rangle$. We take the derivative with respect to $u^{*} (t)$ to find the optimal mode $u(t)$:
\begin{align}
    &\frac{\partial \mathcal{L}[u]}{\partial u^{*}} 
    = \alpha_I^{*} (t) \left( 4 \operatorname{Re} \int_0^T \dd t \, u(t) \alpha_I(t) \right) \\
    &\quad - \lambda \int_{0}^T \dd t' \, \fluctuation (t, t') u(t') 
    - \lambda \int_{0}^T \dd t' \, B^\dagger (t, t') u^{*} (t') \\
    &\quad - \lambda \int_{0}^T \dd t' \, B^{*} (t, t') u^{*}(t') = 0 \; .
\end{align}
We define $C \equiv 4 \operatorname{Re} \int_0^\infty \dd t \, u(t) \alpha_I(t) \in \mathbb{R}$. This equation for $u(t)$ can be solved by decomposing all the objects in real and imaginary part, e.g. $\fluctuation (t, t') = \operatorname{Re} \fluctuation(t, t') + i \operatorname{Im} \fluctuation(t, t')$, $B (t, t') = \operatorname{Re} B(t, t') + i \operatorname{Im} B(t, t')$, $u(t) = u_R(t) + i u_I(t)$, and $\alpha_I(t) = \operatorname{Re} \alpha_I(t) + i \operatorname{Im} \alpha_I(t)$, obtaining the following equations
\begin{equation}
    \begin{aligned}
    \operatorname{Im} \alpha_I 
        &- \operatorname{Im} \fluctuation u_R - \operatorname{Re} \fluctuation u_I 
        +  (\operatorname{Im} B)^\dagger u_R + (\operatorname{Re} B)^\dagger u_I \\
        & + (\operatorname{Im} B)^{*} u_R + (\operatorname{Re} B)^{*} u_I 
        = 0 \; , \\[6pt]
    \operatorname{Re} \alpha_I 
        &- \operatorname{Re} \fluctuation u_R + \operatorname{Im} \fluctuation u_I 
        - (\operatorname{Re} B)^\dagger u_R + (\operatorname{Im} B)^\dagger u_I \\
        &- (\operatorname{Re} B)^{*} u_R + (\operatorname{Im} B)^{*} u_I 
        = 0 \; ,
    \end{aligned}
\end{equation}
where we have simplified the notation for the integrals to a product, and we have absorbed $\lambda$ and $C$ in the definition of $u(t)$, which need not to be normalized at this stage. This is a system of linear equations with two unknown functions $u_R(t)$ and $u_I(t)$, which we can write as following
\begin{equation}
    \begin{aligned}
    &\operatorname{Im} \alpha_I + [-\operatorname{Im} \fluctuation + (\operatorname{Im} B)^\dagger + (\operatorname{Im} B)^*] u_R \\
    &\quad + [-\operatorname{Re} \fluctuation + (\operatorname{Re} B)^\dagger + (\operatorname{Re} B)^*] u_I = 0 \; , \\
    &\operatorname{Re} \alpha_I + [-\operatorname{Re} \fluctuation - (\operatorname{Re} B)^\dagger - (\operatorname{Re} B)^*] u_R \\
    &\quad + [\operatorname{Im} \fluctuation \fluctuation + (\operatorname{Im} B)^\dagger + (\operatorname{Im} B)^*] u_I = 0 \; ,
    \end{aligned}
\end{equation}
We define
\begin{equation}
    \begin{aligned}
    X_R &\equiv -\operatorname{Im} \fluctuation + (\operatorname{Im} B)^\dagger + (\operatorname{Im} B)^* \; , \\
    X_I &\equiv -\operatorname{Re} \fluctuation + (\operatorname{Re} B)^\dagger + (\operatorname{Re} B)^* \; , \\
    Z_R &\equiv -\operatorname{Re} \fluctuation - (\operatorname{Re} B)^\dagger - (\operatorname{Re} B)^* \; , \\
    Z_I &\equiv \operatorname{Im} \fluctuation + (\operatorname{Im} B)^\dagger + (\operatorname{Im} B)^* \; ,
    \end{aligned}
\end{equation}
and write the solution as
\begin{equation}
    \begin{aligned}
    &u_R = \left( Z_R - Z_I X_I^{-1} X_R \right)^{-1} \left( Z_I X_I^{-1} \operatorname{Im} \alpha_I - \operatorname{Re} \alpha_I \right) \; , \\
    &u_I = \left( Z_R X_R^{-1} + Z_I \right)^{-1} \left( \operatorname{Re} \alpha_I - Z_R X_R^{-1} \operatorname{Im} \alpha_I \right) \; .
    \end{aligned}
\end{equation}
which concludes the derivation of the optimal mode.


\section{Intuition for why the max-SNR mode has almost no photons in it}
\label{sec:interpretation_mode}
In this appendix, we discuss the interpretation of the maximum-SNR mode $w_{\text{SNR}}(t)$, and how it can be informationally optimal while containing a vanishingly small fraction of the total photons. This section is not a derivation of the scaling law for the photon number in Fig.~\eqref{fig:SNR_comparison}, and the results we derive rely heavily on assumptions that are not exactly satisfied by the superradiant system. Nevertheless, the toy models of this section clarify how it is plausible that the number of photons in $w_\text{SNR}(t)$ scales sublinearly in the number of spins. We have not been able to find an analytical derivation of the scaling law that we empirically present in Fig.~\eqref{fig:SNR_comparison}.

\textit{Generic conditions for having a max-SNR mode with almost no power and no signal}. In Eq.~\eqref{eq:homodyne_current}, we have defined a homodyne current operator, from which we derive the signal $\alpha_I(t)$, and the mode of the output radiation which maximizes the signal, i.e., $w_{\text{sig}}(t)$ defined in Eq.~\eqref{eq:most_sensitive}. A trivial, uncorrelated, white noise would imply that the best mode to filter in the homodyne detection is the (conjugated and normalized) signal mode itself. However, in our scenario, the non-trivial structure of the noise means that the optimal mode to measure is not simply the signal noise, as we see in Fig.~\ref{fig:three_modes_comparison}. 

\begin{figure}[tbp]
  \centering
  \includegraphics[width=0.45\textwidth]{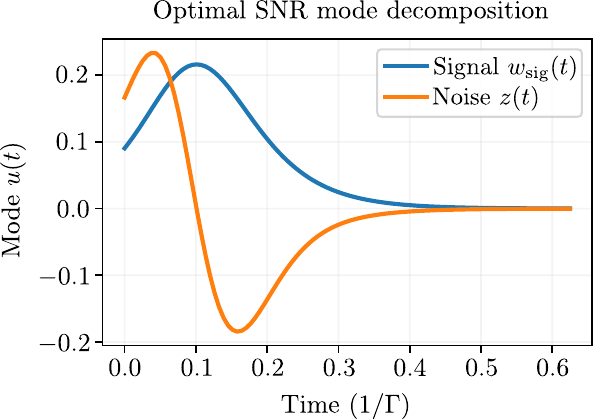} 
  \caption{\justifying Decomposition of the maximum SNR mode in Eq.~\eqref{eq:snr_optimizer} in a signal mode (Eq.~\eqref{eq:most_sensitive}), and a zero signal mode, for a small number $N=16$.}
\label{fig:decomposition_modes}
\end{figure}

To provide some intuition into the structure of the optimal max-SNR mode, we decompose it into a sum of two orthogonal terms: the first term involves the max-signal mode $w_{\text{sig}}(t)$, while the second term involves an orthogonal normalized mode $z(t)$.  For some $0 < s < 1$, we have:  
\begin{equation}
    w_{\text{SNR}}(t) = \sqrt{1-s} \, w_{\text{sig}}(t) + \sqrt{s} \, z(t) \; .
    \label{eq:combination}
\end{equation}
Note that any mode orthogonal to $w_{\text{sig}}(t)$ has zero signal, hence $z(t)$ has no signal (it is pure noise). See Fig.~\eqref{fig:decomposition_modes} for an example of decomposition. 

We now derive a formula for the coefficient $s$, which controls how much signal there is in the max-SNR mode. We indicate the noise (the fluctuations of the homodyne current) of the signal mode $w_{\text{sig}}(t)$ with $f_w$, while we indicate with $f_z$ the noise associated to $z(t)$.  We write the covariance between these fluctuations as $\corrmodes \sqrt{ f_w f_z}$, where $-1 \leq \corrmodes \leq 1$ is the correlation constant.  
The SNR can be written as a function of $s, f_w, f_z$ and $\corrmodes$; up to a prefactor, it is:
\begin{equation}
    {\text{SNR}} \propto \frac{1-s}{(1 - s) f_w + s f_z + 2 \corrmodes \sqrt{s (1-s) f_w f_z}} \; .
\end{equation}
%
By maximizing this expression over $s$ we get
\begin{equation}
    {\text{SNR}} \propto \frac{1}{f_w (1-\corrmodes^2)} \; .
\end{equation}
with the optimal value of $s^{*}$ given by
\begin{equation}
    s^{*} \equiv \frac{\corrmodes^2}{\frac{f_z}{f_w} + \corrmodes^2}.
\end{equation}
The above expression has an interesting limit that is of relevance to our system:  when the noise $f_z \ll f_w$ and for any non-zero correlation constant $\corrmodes$, the mode $w_\text{SNR}$ becomes very close to the zero-signal mode $z(t)$, since $s^{*}$ approaches one. 

We can try to use these generic results to understand why in our superradiant setup, the max-SNR mode has a vanishing fraction of the total photon numbers.  We can do this by using the above expressions and making three mild assumptions:  
\begin{enumerate}
    \item   The correlation constant $\corrmodes$ tends to a non-zero constant in the large-$N$ limit.
    \item   The signal mode fluctuations scale like $N$ in the large-$N$ limit, i.e.~$f_w = \mathcal{O}(N)$.
    \item  The fluctuations in the no-signal mode $z(t)$ grow sublinearly with $N$ in the large $N$ limit, implying $f_z/f_w \rightarrow 0$ for $N\rightarrow\infty$. 
\end{enumerate}
Note that point (2) above is rigorously true for out setup.  

Given these conditions, the variance of the max-SNR mode reduces to:
\begin{equation}
    f_{\text{SNR}} \equiv \frac{\frac{f_z}{f_w} \left( 1 - \corrmodes^2 \right)}{\frac{f_z}{f_w} + \corrmodes^2} f_w \simeq f_z \frac{1 - \corrmodes^2}{\corrmodes^2} \; ,
\end{equation}
where we have simplified $\frac{f_z}{f_w}$ because of assumption $3)$. Assumptions $2)$ and $3)$ now imply that the noise in the max-SNR mode will grow sublinearly with $N$, in the same manner that the noise $f_z$ grows with $N$.  

The upshot of this analysis is to provide an intuitive way of understanding how one could obtain a max-SNR mode that has almost no photons in it.  We need the existence of a no-signal mode that has very few photons in it, but whose noise is correlated with the max-signal mode.  In this case, the above optimization shows that the optimal strategy is to {\it primarily} measure the few-photon, no-signal mode, with just a tiny bit of the signal mode thrown in.  

While this discussion provides some intuition, it of course leaves many things unanswered (e.g.~why does there exist a low-photon number, no-signal mode whose fluctuations are correlated with the max-signal mode?).  

\section{Computation of the quantum states}
\label{sec:cascaded_quantum_state}
Naively, the computation of non-linear functions of the state of the radiation, like the Fisher information, and the quantum Fisher information require the simulation of the state of the radiation itself~\cite{khan_tensor_2025}. This can be accomplished for a handful of modes through the cascaded master equation formalism~\cite{kiilerich_input-output_2019, kiilerich_quantum_2020, combes_slh_2017, carmichael_quantum_1993}.

\begin{figure*}[htbp]
  \centering
  \includegraphics[width=\textwidth]{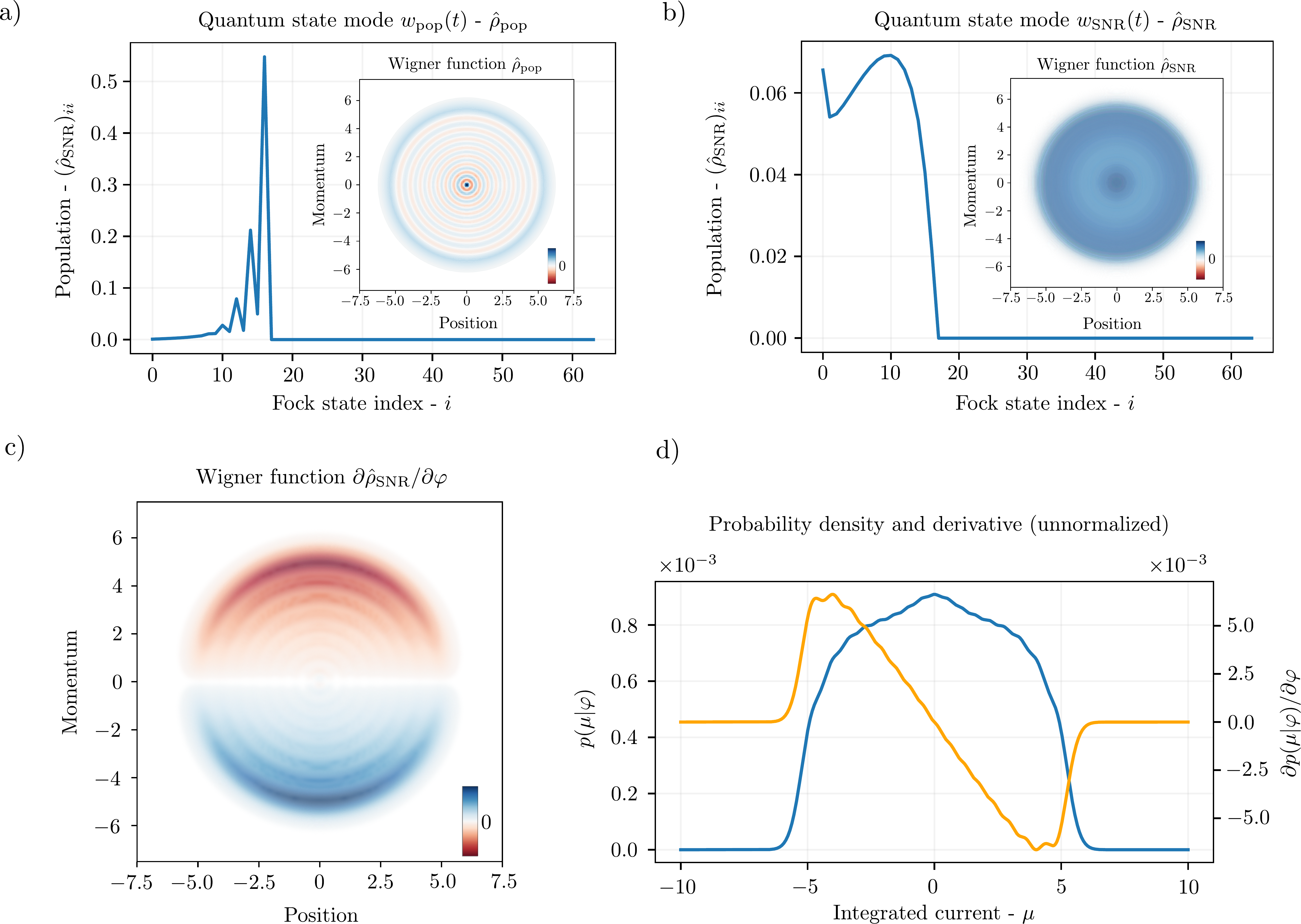} 
  \caption{\justifying a) The picture represents the population of the most populated mode $w_{\text{pop}}(t)$ for $N=16$, the inset is its Wigner function, see Eq.~\ref{eq:most_populated_mode}. b) Population and Wigner function of the mode that optimizes the SNR, see Eq.~\eqref{eq:snr_optimizer}. Contrary to the $w_{\text{pop}}(t)$, $w_\text{SNR}(t)$ has no Wigner negativity. We observe that the uncertainty in the number of photons in the mode $w_{\text{SNR}}(t)$ is larger then the uncertainty for the most populated mode. c) Wigner function of the derivative of the state of the optimal mode (used to compute the QFI). d) Probability density (and it's derivative) of the integrated homodyne current. The density function is not normalized for better visualization in comparison with its derivative.}
\label{fig:cavity_state_comparison}
\end{figure*}

In Fig.~\ref{fig:cavity_state_comparison} we compare the states for the most populated mode in Eq.~\eqref{eq:most_populated_mode}, called $\hat{\rho}_{\text{pop}}$, and for the mode that maximizes the SNR in Eq.~\eqref{eq:snr_optimizer}, called $\hat{\rho}_{\text{SNR}}$. The fact that both these states are mixed is an indication of the entanglements the modes. Because of the $U(1)$ symmetry we can say a priori that the density matrix of the quantum state of the light has zero out-of-diagonal elements. 

From the state of the mode it is possible to compute two information quantities that we compare with the QFI of the spin ensemble and with the SNR obtained from the homodyne measurement on the radiation. We compute the Fisher information associated to the measurement of the integrated homodyne current, i.e. the measurement of the observable $\hat{M}_u$ in Eq.~\eqref{eq:observable}. This precision is achievable with an optimized estimator, that goes beyond Eq.~\eqref{eq:estimator_varphi}. The second information quantity we compute is the QFI of the mode, which is the ultimate precision limit in the estimation of $\parameter$ associated with the most general measurement on the reduced state of the mode only. We evaluate these quantities for the encoding of the rotation angle in Eq.~\eqref{eq:initial_state_separable}, where the QFI of the spin ensemble is the standard quantum limit, i.e. $\mathcal{F}(\ket{\psi_\parameter^S}) = N$, for small values of the number of spins $N$. We report these results in Table~\ref{tab:qfi_fi_results}. From this table we conclude that were we able to extract the full QFI from the state of the mode $\hat{\rho}_{\text{SNR}}$, the improvement over the linear estimation would be marginal. Remember that the mode in Eq.~\eqref{eq:snr_optimizer} is optimized to maximize the SNR and not the QFI. We can't exclude the existence of an optimal mode for $F(\hat{\rho}_{\text{SNR}})$ of $\mathcal{F}(\hat{\rho}_{\text{SNR}})$ that achieves the full QFI of the spin ensemble. The computation of these modes, which we think can only be done numerically goes beyond the scope of this paper.
\begin{table}[ht]
\centering
\begin{tabular}{@{}ccc@{}}
\toprule 
\textbf{Num. spins} ($N$) & $\mathcal{F}(\hat{\rho}_{\text{SNR}})/N$ & $F(\hat{\rho}_{\text{SNR}})/N$ \\ \midrule
4 & 0.77 & 0.53   \\
8 & 0.54  & 0.45  \\
16 & 0.47  & 0.46 \\
\bottomrule
\end{tabular}
\caption{\justifying QFI and FI for the mode $w_{\text{SNR}}(t)$ in Eq.~\eqref{eq:snr_optimizer}, for small values of the number of spins in the ensemble. This information quantities are to be compared with the SNR achievable by the linear estimator, see Fig.~\ref{fig:three_modes_comparison}.}
\label{tab:qfi_fi_results}
\end{table}

\section{Optimality of the homodyne and photon counting measurements}
\label{sec:optimality_unravelling}
In this appendix, we prove the optimality of homodyne detection and photon counting for the estimation of the rotation angle imprinted on the initial state of the spin ensemble in Eq.~\eqref{eq:initial_state_separable}. This proof is valid if the cavity is perfectly in resonance with the spin ensemble, and if the spins are initially fully excited.

\textit{Equivalence with a continuous encoding problem}. Recently, there has been progress in the quest of defining the maximum achievable precision in the estimation of continuously encoded parameters, together with the quest for finding the measurement that achieves this maximum~\cite{catana_fisher_2015, khan_tensor_2025, yang_quantum_2025}. We show in this section how the initial-value problem codified by Eq.~\eqref{eq:initial_state_separable} can be transformed into a continuous encoding scenario with a dissipator that depends on $\parameter$, the parameter to be estimated. This allows us to exploit the results of Ref.~\onlinecite{mattes_designing_2025} and to prove that the Fisher information of the continuous measurements (photon counting and homodyne detection) achieves exactly the quantum Fisher information in Eq.~\eqref{eq:standard_quantum_limit}. The estimator for $\parameter$, needed to saturate the QFI, involves in principle all complex higher order temporal correlation functions of the homodyne current. This proof does not extend to the case where there is detuning between the cavity and the spins, or to the use of twist-untwist encoding. Therefore it does not extend to the achievability of the Heisenberg scaling. It is surprising that the QFI of the entangled and multi-modal state of the light is exactly achievable with local measurements on the output field.


\textit{New reference frame for the spin ensemble}. In the following, we give an equivalent description of the state of the light emitted by the spins in superradiance, which is based on a continuous encoding scenario instead of the initial-value problem. We indicate with $\hat \rho_\parameter$ the pure state in Eq.~\eqref{eq:initial_state_separable}. At time $t = 0$ the initial state of the spin ensemble and the waveguide is $\hat{\rho}_\parameter \otimes \ket{0} \! \bra{0}$, with the waveguide being in the vacuum state. We call $\hat{U}_{SR}^{(0 \rightarrow t)}$ the joint unitary evolution of the ensemble of spins and the infinitely many modes of the waveguide, such that we can write the state of the radiation field at time $t$ by tracing out the spins:
\begin{equation}
    \hat{\rho}_R^\parameter (t) \equiv \trsquareS{\hat{U}_{SR}^{(0 \rightarrow t)} \left( \hat{\rho}_\parameter \otimes \ket{0} \! \bra{0} \right) \hat{U}_{SR}^{(0 \rightarrow t) \dagger}} \; .
    \label{eq:original_unitary_rotation}
\end{equation}
We apply a change of reference frame to the ensemble of spins, by acting with the inverse of the unitary in Eq.~\eqref{eq:initial_state_separable}, which is just a rotation around the $y$-axis, that undoes the encoding of the spins:
\begin{equation}
\begin{split}
    \hat{\rho}_R^\parameter (t) \equiv \text{Tr}_S \Big[ & \hat{U}_{-\parameter} \hat{U}_{SR}^{(0 \rightarrow t)} \hat{U}_{-\parameter}^\dagger \hat{U}_{-\parameter} \left( \rho_\parameter \otimes \ket{0} \! \bra{0} \right) \\
    & \times \hat{U}_{-\parameter}^\dagger \hat{U}_{-\parameter} \hat{U}_{SR}^{\dagger (0 \rightarrow t)} \hat{U}_{-\parameter}^\dagger \Big] \; .
\end{split}
\end{equation}
Inserting the unitary $\hat{U}_{-\parameter}$, which is acting only on the spins, leaves the trace unchanged. The joint initial state of the spins and the waveguide becomes independent of $\parameter$ in the new frame:
\begin{equation}
    \hat{U}_{-\parameter} \left( \hat{\rho}_\parameter \otimes \ket{0} \! \bra{0} \right)  \hat{U}_{-\parameter}^\dagger = \hat{\rho} \otimes \ket{0} \! \bra{0} \; ,
\end{equation}
where $\hat{\rho} \equiv \ket{j} \! \bra{j}$ is the fully excited state of the spins.

As for the action of the change of reference on the interaction unitary $\hat{U}_{SR}^{(0 \rightarrow t)}$, we can start from the explicit definition of $\hat{U}_{SR}^{(0 \rightarrow t)}$, which involves the spins and the radiation modes of the waveguide while the cavity has been adiabatically eliminated. The definition of this unitary is
\begin{equation}
    \hat{U}_{SR}^{(0 \rightarrow t)} = \exp \left[ - i \left( \hat{H}_R + \hat{H}_S + \sum_{k} (g_k \hat{J}_+ \hat{b}_k + g_k^{*} \hat{J}_{-} \hat{b}_k^\dagger) \right) \right] \; ,
    \label{eq:original_interaction}
\end{equation}
where $\hat{H}_R$ is the free Hamiltonian of the radiation, $\hat{H}_S$ is the Hamiltonian of the spins, and the interaction term features the lowering and raising operators of the spin ensemble coupled to the bosonic operators of the waveguide modes in the momentum basis, $\hat{b}_k$ and $\hat{b}_k^\dagger$. The action of $\hat{U}_{-\parameter}$ leaves $\hat{H}_R$ unchanged, while the free Hamiltonian of the spins become $\hat{H}_S \rightarrow \hat{U}_{-\parameter} \hat{H}_S \hat{U}_{-\parameter}^\dagger$. Similarly the angular momentum operators in the interaction term become $\hat{J}_{-} \rightarrow \hat{U}_{-\parameter} \hat{J}_{-} \hat{U}_{-\parameter}^\dagger$ and  $\hat{J}_{+} \rightarrow \hat{U}_{-\parameter} \hat{J}_{+} \hat{U}_{-\parameter}^\dagger$. We now define the unitary operator for the superradiant evolution in the new reference frame as follows.
\begin{equation}
    \hat{U}_{SR, \parameter}^{(0 \rightarrow t)} \equiv \hat{U}_{-\parameter} \hat{U}_{SR}^{(0 \rightarrow t)} \hat{U}_{-\parameter}^\dagger \; ,
    \label{eq:rotated_unitary_interaction}
\end{equation}
so that the state of the light at time $t$ is described by tracing out the spin subject to the $\parameter$-dependent evolution $\hat{U}_{SR, \parameter}^{(0 \rightarrow t)}$, i.e.
\begin{equation}
     \hat{\rho}_R^\parameter (t) = \trsquareS{\hat{U}_{SR, \parameter}^{(0 \rightarrow t)} \left( \hat{\rho} \otimes \ket{0} \! \bra{0} \right) \hat{U}_{SR, \parameter}^{(0 \rightarrow t) \dagger}} \; ,
     \label{eq:rotated_unitary_photons}
\end{equation}
In order to describe the reduced dynamics of the spins only, we can trace out the degrees of freedom of the radiation field to obtain the master equation for the spins. For the Hamiltonian part we can assume $\hat{H}_S=0$, i.e. we choose the original frame of reference to be rotating with frequency of the spins. Starting from Eq.~\eqref{eq:rotated_unitary_interaction}, we write explicitly the form of the interaction $\hat{U}_{SR, \parameter}^{(0 \rightarrow t)}$ as
\begin{multline}
    \hat{U}_{SR, \parameter}^{(0\to t)}=\exp\!\Big[-i\Big(\hat{H}_R+\sum_k (g_k\hat{U}_{-\parameter}\hat{J}_+\hat{U}^\dagger_{-\parameter}\hat{b}_k \\ + g_k^*\hat{U}_{-\parameter}\hat{J}_-\hat{U}^\dagger_{-\parameter}\hat{b}_k^\dagger)\Big)\Big] \; .
    \label{eq:rotated_interaction}
\end{multline}
Comparing this expression with Eq.~\eqref{eq:original_interaction} we see that the superradiant decay of the ensemble can be described with the superoperator $\mathcal{D}[\hat{U}_{-\parameter} \hat{J}_{-} \hat{U}_{-\parameter}^\dagger]$.
To be explicit, in the new frame the dynamics of the spin ensemble is describe by the $\parameter$ dependent Lindbladian:
\begin{equation}
    \partial_t \hat \rho(t) = \Gamma \mathcal{D}[\hat{U}_{-\parameter} \hat{J}_{-} \hat{U}_{-\parameter}^\dagger] \hat \rho(t).
\end{equation}
This describes a  superradiant decay process along an axis tilted from $z$ by an angle $\parameter$. At the same time, the initial state of the spins is $\hat{\rho}(0) \equiv \singleket{j} \! \singlebra{j}$.

Note that a byproduct of this transformation is an alternate, but equivalent way to look at the main results of our work.  It also describes the measurement physics of a particular class of continuous metrology setups where the parameter of interest modulates a collective spin dissipator.  Our results our also striking in this setting: we show a simple way of achieving Heisenberg-limited scaling in $N$ using a simple homodyne measurement of the output field.  This should be contrasted against schemes where achieving optimal scalings require a second physical system to processes the output field before measurement, e.g.~a coherent quantum absorber~\cite{yang_efficient_2023, girotti_estimating_2025}.

\textit{Information analysis of the continuous encoding scenario}. In the previous paragraphs, we have constructed a conceptually different encoding process for the rotation angle $\parameter$, which gives us an alternative and completely equivalent way to compute the state of the light during the superradiant decay. We define two scenarios, called scenario I and scenario II. The first one is an initial-value problem, where the rotation angle is encoded in the initial state of the spin ensemble, and the decay operator on the spins is $\mathcal{D}[\hat{J}_{-}]$. This is the scenario of the main text. Scenario II is a continuous encoding scheme, where the decay operator is $\mathcal{D}[\hat{U}_{-\parameter} \hat{J}_{-} \hat{U}_{-\parameter}^\dagger]$, and was introduced above. We are interested in working in the limit $t\rightarrow\infty$, where the state of the radiation and the spins is separable. While the state of the radiation is the same in both scenarios, the state of the spins is different. In scenario I, the spins are in the lower energy state $\ket{ -j}$, while in scenario II, the final state of the spins is rotated $\hat{U}_\parameter \ket{-j}$. This makes a difference for the computation of the total asymptotic quantum Fisher information of the joint system. In scenario I, the spins have no information on $\parameter$ because their state is completely decayed, while the radiation contains the full quantum Fisher information of the initial state of the ensemble of spins, i.e. $\mathcal{F}_\parameter(\hat{\rho}^\parameter_R (t\rightarrow\infty)) = N$. If we consider instead scenario II, we find that the spins have been encoded with the parameter $\parameter$, so that the final QFI is $\mathcal{F}_\parameter(\hat{U}_\parameter \ket{-j}) + \mathcal{F}_\parameter(\hat{\rho}^\parameter_R (t\rightarrow\infty)) = 2N$, i.e., twice that of scenario I. This also underlines how these two scenario are not equivalent in terms of the metrological analysis, only the state of the light produced is the same.

\textit{Continuous measurements}. We now consider the use of continuous measurements, measuring the radiation as it is generated by the decaying spins. We use the term \text{trajectory} to indicate the full list of observed results of the continuous measurement. We indicate with $F^{h, c}(p_\text{traj})$ the classical Fisher information we can extract from the full trajectory of outcomes from these measurements, where $p_\text{traj}$ is the probability distribution for the trajectories.  We use the notation $F^c$ for the CFI of photon counting, while $F^h$ is the CFI of the homodyne detection. 

In general, after the continuous measurement, the spin ensemble will be in a conditional state that depends on the trajectory, which we indicate with $\hat{\rho}_\parameter^\text{traj}$. A final measurement is performed on this state of the spins, with the optimal measurement saturating the QFI. From Ref.~\cite{albarelli_ultimate_2017}, we write the total Fisher information extracted from the continuous measurement as
\begin{equation}
    F^{h, c} = F^{h, c}(p_\text{traj}) + \mathbb{E}_{\text{traj}} [\mathcal{F}_\parameter (\hat{\rho}_\parameter^\text{traj})] \; .
    \label{eq:total_qfi_continuous_measurement}
\end{equation}
In scenario I (initial-value encoding), the final state of the spins long after the decay is independent of the trajectory and of $\parameter$, so that we have
\begin{equation}
     F_{I}^{h, c} = F^{h, c}_{I}(p_\text{traj}) \; .
\end{equation}
In scenario II (continuous encoding), the final state of the spins is also independent of the trajectory and we have
\begin{align}
    F_{II}^{h, c} &= F^{h, c}_{II}(p_\text{traj}) + \mathcal{F}_\parameter(\hat{U}_\parameter \ket{-j}) \\
             &= F^{h, c}(p_\text{traj}) + N \; .
\end{align}
We will prove in the next paragraph that homodyne detection and photon counting both saturate the joint QFI of the spins and the radiation field, so that we can write $F_{II}^{h, c}(p_\text{traj}) + N = 2 N$, which implies $F_{II}^{h, c}(p_\text{traj}) = N$. The final observation to connect this result to scenario I is that $F_{II}^{h, c}(p_\text{traj})$ depends only on the state of the radiation, which is the same in scenario I and II, which implies $F_{I}^{h, c}(p_\text{traj}) = F_{II}^{h, c}(p_\text{traj})$, so that we have
\begin{equation}
    F_{I}^{h, c}(p_\text{traj}) = \mathcal{F} (\ket{\psi^S_\parameter})\; ,
\end{equation}
since $\mathcal{F} (\ket{\psi^S_\parameter}) = N$. The continuous measurement on the state of the radiation field is sufficient to achieve the QFI in the initial-value problem, when the full statistics deriving from the outcome trajectories is accounted for. In the rest of this section, we prove that the continuous measurements in scenario II are optimal. We start by computing a simple formula for the global QFI of the spins and the radiation field, and we conclude by proving that $F_{II}^{c, h}$ achieves it for both measurements.

\textit{Continuous measurements from a collisional model}. Within a collisional model description of the environment of the spins~\cite{ciccarello_quantum_2022}, we compute the joint evolution of the system and the environment by discretizing the time in small steps $\Delta t$, and introducing a new instance of the Hilbert space of a single mode of the radiation field for each time step. The same reduced dynamics on the system of spins can be described by different Kraus operators according to the continuous measurement performed on the radiation modes (photon counting or homodyne detection). The outcome of the measurement on the environment at each time step is $i_j$, with $j$ being the index of the time step. These outcomes are also the indices of the Kraus operators for each instantaneous measurement, and they can be a continuous index, i.e., $J_j \equiv i_j \in \mathbb{R}$, or a binary value, i.e. $i_j \in \lbrace 0, 1\rbrace$. We now define the Kraus operators for photon counting and homodyne detection, for a single timestep, so that we can suppress the subscript $j$ for simplicity. Let us define the rotated jump operator $\hat{L}_\parameter \equiv \sqrt{\Gamma} \hat{U}_{-\parameter} \hat{J}_{-} \hat{U}_{-\parameter}^\dagger$, then the Kraus operators corresponding to photon counting in scenario II are
\begin{equation}
    \hat{K}_0^\parameter \equiv e^{-\frac{1}{2} \hat{L}_\parameter^\dagger \hat{L}_\parameter \Delta t} \; , \quad \hat{K}_1^\parameter \equiv \sqrt{\Delta t} \hat{L}_\parameter \; ,
    \label{eq:kraus_photon_counting}
\end{equation}
these two operators correspond to no-photons detected ($\hat{K}_0^\parameter$), and a single photon collected ($\hat{K}_1^\parameter$). The Kraus operators satisfy $\hat{K}_0^{\parameter \dagger} \hat{K}_0^{\parameter} + \hat{K}_1^{\parameter \dagger} \hat{K}_1^\parameter = \mathds{1}$ to leading order in $\Delta t$. We now discuss the form of the Kraus operators for homodyne detection. In order to describe homodyne detection on the quadrature $\hat{L}_\parameter e^{i \theta} + \hat{L}_\parameter^\dagger e^{-i \theta}$ we need a continuous set of Kraus operators, indexed by the value of the homodyne current measured at each time step, i.e.
\begin{equation}
    \hat{K}_J^\parameter \equiv 1 - \frac{\hat{L}_\parameter^{\dagger} \hat{L}_\parameter}{2} \Delta t + \hat{L}_\parameter e^{i \theta} J \Delta t \; ,
\end{equation}
which, assuming $\Delta t \rightarrow 0$, we can write in exponential form as
\begin{equation}
    \hat{K}_J^\parameter \simeq \exp \left( - \frac{\hat{L}_\parameter^\dagger \hat{L}_\parameter}{2} \Delta t + \hat{L}_\parameter e^{i \theta} J \Delta t \right) \; ,
    \label{eq:kraus_homodyne}
\end{equation}
which will be useful in simplifying the derivatives in the rest of this section. The Kraus operators for homodyne detection satisfy
\begin{equation}
    \frac{\Delta t}{2 \pi} \int_{-\infty}^{+\infty} \dd J \, e^{-\frac{\Delta t}{2} J^2} \hat{K}_i^{\parameter \dagger} \hat{K}_i^\parameter = \mathds{1} + \mathcal{O}(\Delta t^2) \; .
\end{equation}
The weight $\dd J \, e^{-\frac{\Delta t}{2} J^2}$ is called the ``ostensive probability'' and it is needed to properly normalize the Kraus operators with continuous-measurement outcomes~\cite{wiseman_quantum_2009}. In this section, when otherwise not specified we will always assume $\theta=0$. We indicate with $(i_j)_{j=1}^M$ the tuple containing the full set of outcomes, i.e., the trajectory. Given a certain trajectory, the unnormalized state of the ensemble of spins is
\begin{equation}
    \ket{\tilde{\psi}_{\text{traj}} (\parameter)} \equiv \hat{K}^\parameter_{i_1} \hat{K}^\parameter_{i_2} \hat{K}^\parameter_{i_3} \dots \hat{K}^\parameter_{i_M} \ket{\psi} \; ,
    \label{eq:unnormalized_state}
\end{equation}
while the state of the copy of the environment at the time step $j$ is $\ket{i_j}$, so that $\ket{i_1, \dots, i_M}$ is the full state of the environment. Summing over all possible trajectories gives the full entangled state of the spins and the radiation:
\begin{equation}
    \ket{\Psi_\parameter (t)} \equiv \sum_{( i_j )_{j=1}^M} \hat{K}_{i_1}^\parameter \hat{K}_{i_2}^\parameter \hat{K}_{i_3}^\parameter \dots \hat{K}_{i_M}^\parameter \ket{\psi} \otimes \ket{i_1, \dots, i_M} \; .
\end{equation}

\textit{Expression for the joint QFI}. The quantum Fisher information of the joint state $\ket{\Psi_\parameter (t)}$ is
\begin{equation}
    \mathcal{F}_\parameter (\ket{\Psi_\parameter (t)}) = 4 \left( \braket{\partial _\parameter\Psi_\parameter (t)| \partial_\parameter \Psi_\parameter (t)} + \braket{\Psi_\parameter (t) | \partial_\parameter \Psi_\parameter (t)}^2 \right) \; ,
    \label{eq:quantum_fisher_global}
\end{equation}
where we have used the notation $\ket{\partial_\parameter \Psi_\parameter (t)} \equiv \partial \ket{\Psi_\parameter (t)} / \partial \parameter$. We first write the expression for $\ket{\partial_\parameter \Psi_\parameter (t)}$ in terms of the Kraus operators
\begin{align}
\ket{\partial_\parameter \Psi_\parameter (t)} 
    &= \sum_{j=1}^M \sum_{( i_l )_{l=1}^M} 
        \hat{K}_{i_1}^\parameter \hat{K}_{i_2}^\parameter \dots 
        \hat{K}^\parameter_{i_{j-1}} \left( \partial_\parameter
        \hat{K}_{i_j}^\parameter \right) \hat{K}^\parameter_{i_{j+1}}
        \nonumber \\[-2pt]
    &\qquad\qquad
        \dots \hat{K}_{i_M}^\parameter 
        \ket{\psi} \otimes \ket{i_1, i_2, \dots, i_M}
        \\[6pt]
    &= \sum_{( i_l )_{l=1}^M} \sum_{j=1}^M  
        \hat{K}_{i_1}^\parameter \hat{K}_{i_2}^\parameter \dots 
        \hat{K}^\parameter_{i_{j-1}} \left( \partial_\parameter
        \hat{K}_{i_j}^\parameter \right) \hat{K}^\parameter_{i_{j+1}}
        \nonumber \\[-2pt]
    &\qquad\qquad
        \dots \hat{K}_{i_M}^\parameter 
        \ket{\psi} \otimes \ket{i_1, i_2, \dots, i_M}
        \; ,
\end{align}
where in the second equation we have swapped the summations. We can compute the norm of the derivative $\ket{\partial_\parameter \Psi_\parameter (t)}$ by observing that states of the environment corresponding to different trajectories are orthogonal to each other, so that the total norm of the derivative is the sum of the norms of the derivatives of the unnormalized states of the spins:
\begin{multline}
    \braket{\partial _\parameter\Psi_\parameter (t)| \partial_\parameter \Psi_\parameter (t)} \\ = 
    \sum_{( i_l )_{l=1}^M} \left\| \sum_{j=1}^M \hat{K}^\parameter_{i_1} \hat{K}^\parameter_{i_2} \cdots \left( \partial_\parameter \hat{K}^\parameter_{i_j} \right) \cdots \hat{K}^\parameter_{i_M} \ket{\psi} \right\|^2 \; .
    \label{eq:first_term_QFI}
\end{multline}
We now consider the second term of the QFI in Eq.~\eqref{eq:quantum_fisher_global}, and we write explicitly the expression for the scalar product of $\ket{\Psi_\parameter (t)}$ and its derivative, by introducing the Kraus operators and the summation over the trajectories:
\begin{align}
    \langle \Psi_\parameter (t) &| \partial_\parameter \Psi_\parameter (t) \rangle \nonumber = \sum_{j=1}^M \sum_{( i_l )_{l=1}^j} \langle\psi | \hat{K}^{\parameter \dagger}_{i_1} \dots \hat{K}^{\parameter \dagger}_{i_{j-1}} \hat{K}^{\parameter \dagger}_{i_j} \nonumber \\
    &\quad \times \left( \partial_\parameter \hat{K}^\parameter_{i_j} \right) \hat{K}^{\parameter \dagger}_{i_{j-1}}\dots \hat{K}^\parameter_{i_1}| \psi\rangle \; .
    \label{eq:second_term_starting_point}  
\end{align}
We will prove in the next paragraphs that this term is zero, both for homodyne detection and for photon counting, so that we can simplify the expression for the QFI to the term in Eq.~\eqref{eq:first_term_QFI} only. The state of the environment and the spins at a certain time is independent of the choice of the measurement, therefore we need to prove the statement only either for photon counting or homodyne detection, that is to say both unravellings of the dynamics with different Kraus operators must give the same results for $ | \Psi_\parameter (t) \rangle$ and its derivatives. Nevertheless in the following we report both proofs, in order to introduce the two different expressions for the Fisher information for the two different continuous measurements.

\textit{Joint QFI of the spins and the radiation for photon counting.} We commute the summation on the time index and on the trajectories, and write two separate families of trajectories, according to the value of $i_j$:
\begin{equation}
\begin{aligned}
\langle \Psi_\parameter (t) &| \partial_\parameter \Psi_\parameter (t) \rangle
    \\ &= \sum_{j=1}^M \sum_{\substack{(i_l)_{l=j}^M \\ i_j = 0}} \braket{\psi | \hat{K}^{\parameter \dagger}_{i_1} \dots \hat{K}^{\parameter \dagger}_{0} \left( \partial_\parameter \hat{K}^\parameter_{0} \right) \dots \hat{K}^\parameter_1| \psi} \\ &\quad + \sum_{j=1}^M \sum_{\substack{(i_l)_{l=j}^M \\ i_j = 1}} \braket{\psi | \hat{K}^{\parameter \dagger}_{i_1} \dots \hat{K}^{\parameter \dagger}_{1} \left( \partial_\parameter \hat{K}^\parameter_{1} \right) \dots \hat{K}^\parameter_{i_1}| \psi}
\end{aligned}
\label{eq:second_term_photon counting}
\end{equation}
We write explicitly the term
\begin{equation}
    \hat{L}_\parameter^\dagger \hat{L}_\parameter = \hat{U}_{-\parameter} \hat{J}_{+} \hat{J}_{-} \hat{U}_{-\parameter}^\dagger = j(j+1) - \hat{U}_{-\parameter} J_z^2 \hat{U}_{-\parameter}^\dagger \; .
\end{equation}
So that we can compute the derivative
\begin{equation}
    \hat{K}_0^{\parameter \dagger} \partial_\parameter \hat{K}^\parameter_0 = -\frac{i}{2} \Gamma \Delta t [\hat{J}_y, \hat{J}_z^2] \Delta t \propto  \lbrace \hat{J}_x, \hat{J}_z \rbrace \; .
    \label{eq:derivative_K_0}
\end{equation}
The first summand of Eq.~\eqref{eq:second_term_photon counting} is proportional to the expectation value of the operator $\lbrace \hat{J}_x, \hat{J}_z \rbrace$ evaluated on the unnormalized state of the spins evolved up to time $t = j\Delta t$, summed over all possible trajectories. Because of the sum over all trajectories we can turn this term into the weighted expectation value over the normalized state conditioned on the trajectory, which is simply the expectation value over the reduced state of the spins. In other words, the first summand evaluated to the integral of $\langle \lbrace \hat{J}_x (t), \hat{J}_z (t) \rbrace \rangle$, where $\hat{J}_x (t)$ and $\hat{J}_z (t)$ are the spin operators in the Heisenberg representation and the expectation value is computed over the initial state of the spins $\ket{j}$. It can be verified numerically that this expectation value is identically zero, because of the rotational symmetry of the initial state around the $z$ axis. We now focus on the second summand of Eq.~\eqref{eq:second_term_photon counting}. We write the innermost product as
\begin{equation}
    \hat{K}_1^{\parameter \dagger} \partial_\parameter \hat{K}^\parameter_1 \propto \hat{J}_{+} [\hat{J}_y, \hat{J}_{-}] = \hat{J}_{+} [\hat{J}_y, \hat{J}_x] \propto \hat{J}_{+} \hat{J}_{z} \; ,
\end{equation}
which, because of rotational symmetry, also has zero expectation value. We conclude that the expression in Eq.~\eqref{eq:second_term_photon counting} it is identically zero, i.e. $\braket{\Psi_\parameter (t) | \partial_\parameter \Psi_\parameter (t)} = 0$. We finally write the expression for the total QFI of the joint system of spins and radiation, valid for every time $t$:
\begin{align}
    \mathcal{F}_\parameter & (\ket{\Psi_\parameter (t)}) = 4 \langle \partial_\parameter \Psi_\parameter (t) | \partial_\parameter \Psi_\parameter (t) \rangle \nonumber \\
    &= 4 \sum_{( i_j )_{j=1}^M} \left\| \sum_{j=1}^M \hat{K}^\parameter_{i_1} \hat{K}^\parameter_{i_2} \cdots \left( \partial_\parameter \hat{K}^\parameter_{i_j} \right) \cdots \hat{K}^\parameter_{i_M} \ket{\psi} \right\|^2 \; .
    \label{eq:fisher_photon_counting}
\end{align}

\textit{Joint QFI of the spins and the radiation for homodyne detection.} Because of the different normalization of the Kraus operators for homodyne detection, the joint state of the ensemble of spins and the radiation is written as
\begin{align}
    & \ket{\Psi(t)} = \left( \frac{\Delta t}{2 \pi} \right)^{\frac{M}{4}} \! \! \! \! \int_{-\infty}^{+\infty} dJ_M e^{-\frac{\Delta t}{4} J_M^2} \cdots \int_{-\infty}^{+\infty} dJ_1 e^{-\frac{\Delta t}{4} J_1^2} \nonumber \\
    & \quad \cdot \hat{K}^\parameter_{J_M} \hat{K}^\parameter_{J_{M-1}} \cdots \hat{K}^\parameter_{J_j} \cdots \hat{K}^\parameter_{J_1} \ket{\psi} \otimes \ket{J_1, J_2, \cdots, J_M} \; .
\end{align}
We evaluate the derivative of this state as a function of the parameter $\parameter$:
\begin{align}
    &\ket{\partial_\parameter \Psi(t)} = \sum_{j=1}^M \left( \frac{\Delta t}{2 \pi} \right)^{\frac{M}{4}} \! \! \! \! \int_{-\infty}^{+\infty} dJ_M e^{-\frac{\Delta t}{4} J_M^2} \cdots \nonumber \\
    &\quad \cdot \hat{K}^\parameter_{J_M} \cdots \left( \partial_\parameter \hat{K}^\parameter_{J_j} \right) \cdots \hat{K}^\parameter_{J_1} \ket{\psi} \otimes \ket{J_1, \cdots, J_M} \; .
\end{align}
Similarly to the photon counting case we write
\begin{align}
    &\braket{\Psi(t) | \partial_\parameter \Psi(t)}
    = \sum_{j=1}^M \left( \frac{\Delta t}{2 \pi} \right)^{\frac{M}{4}} \! \! \! \!
       \int_{-\infty}^{+\infty} dJ_M \, e^{-\frac{\Delta t}{4} J_M^2} \cdots \nonumber \\
    &\quad \int_{-\infty}^{+\infty} dJ_1 \, e^{-\frac{\Delta t}{4} J_1^2}
       \braket{\psi | \hat{K}^{\parameter \dagger}_{J_1} \dots \hat{K}^{\parameter \dagger}_{J_j}
       \left( \partial_\parameter \hat{K}^\parameter_{J_j} \right) \dots \hat{K}^\parameter_{J_1} | \psi} \; .
    \label{eq:second_term_starting_point}
\end{align}
We write the innermost product of operators as
\begin{align}
    \hat{K}^{\parameter \dagger}_{J_j} \partial_\parameter \hat{K}^\parameter_{J_j} &= \frac{\partial}{\partial \parameter} \left( 1 - \frac{\hat{L}_\parameter^\dagger \hat{L}_\parameter}{2} \Delta t + \hat{L}_\parameter J_j \Delta t \right) \\
    &\propto \frac{1}{2} [\hat{J}_y, \hat{J}_{+} \hat{J}_{-}] + [\hat{J}_y, \hat{J}_{-}] \hat{J}_j \\
    &= - \{ \hat{J}_x, \hat{J}_z \} + \hat{J}_z J_j
\end{align}
Similarly as in the photon counting paragraph, the expression in Eq.~\eqref{eq:second_term_starting_point} now contains the expectation value $\langle {\hat{J}_x(t), \hat{J}_z(t)} \rangle$, which is zero, and the expectation value of $\hat{J}_z J_j$, which is not computable from the master equation alone, since it depends on the current $J_t$. We average the quantity $\hat{J}_z J_t$ over the trajectories, so we write
\begin{multline}
    \int_{-\infty}^{+\infty} dJ_j e^{-\frac{\Delta t}{4} J_j^2} \cdots \int_{-\infty}^{+\infty} dJ_1 e^{-\frac{\Delta t}{4} J_1^2} \\ \cdot J_j \braket{\psi | \hat{K}^{\parameter \dagger}_{J_1} \hat{K}^{\parameter \dagger}_{J_2} \dots \hat{K}^{\parameter \dagger}_{J_j} \hat{J}_z \hat{K}^{\parameter}_{J_j} \dots \hat{K}^\parameter_{J_2} \hat{K}^\parameter_{J_1}| \psi} \; .
\label{eq:homodyne_second_term_simplified}
\end{multline}
The braket in this expression is proportional to the expectation value $\langle \hat{J}_z(t) \rangle$, evaluated at time $t=j \Delta t$. We can approximate it with the expectation value at time $t=(j-1) \Delta t$, so that the entire braket is independent on $J_j$. This approximation relies on $\Delta t \rightarrow 0$ and the continuity of the state of the spins in each trajectory. At this point we single out the integral of $J_j$, which is zero: $\int_{-\infty}^{+\infty} \dd J_j e^{-\frac{\Delta t}{4} J_j^2} J_j = 0$, which shows that the expression in Eq.~\eqref{eq:homodyne_second_term_simplified} averages to zero. We conclude that similarly to photon counting the joint quantum Fisher information, in terms of the Kraus operators can be expressed as
\begin{align}
    \mathcal{F}_\parameter & (\ket{\Psi_\parameter (t)}) = 4 \langle \partial_\parameter \Psi_\parameter (t) | \partial_\parameter \Psi_\parameter (t) \rangle \nonumber \\
    &= 4 \left( \frac{\Delta t}{2 \pi}\right)^{\frac{M}{2}} \int_{-\infty}^{+\infty} dJ_M e^{-\frac{\Delta t}{2} J_M^2} \cdots \int_{-\infty}^{+\infty} dJ_1 e^{-\frac{\Delta t}{2} J_1^2} \nonumber \\
    &\quad \cdot \left\| \sum_{j=1}^M \hat{K}^\parameter_{J_M} \cdots \left( \partial_\parameter \hat{K}^\parameter_{J_j} \right) \cdots \hat{K}^\parameter_{J_1} \ket{\psi} \right\|^2
\end{align}

\textit{Fisher information of the continuous measurements}. Let us focus on the Fisher information of the continuous measurement, that we can extract from the observation of the trajectories. The probability of observing a certain trajectory $(i_1, i_2, \cdots, i_M)$ is
\begin{equation}
    p_{\text{traj}}(\parameter) \equiv \norm{\hat{K}^\parameter_{i_1} \hat{K}^\parameter_{i_2} \hat{K}^\parameter_{i_3} \dots \hat{K}^\parameter_{i_M} \ket{\psi}}^2 \; .
    \label{eq:probability_trajectory}
\end{equation}
Having defined the unnormalized state  $\ket{\tilde{\psi}_{\text{traj}} (\parameter)}$ associated with a certain observed trajectory in Eq.~\eqref{eq:unnormalized_state}, we can write $p_{\text{traj}}(\parameter) = \braket{\tilde{\psi}_{\text{traj}} (\parameter) | \tilde{\psi}_{\text{traj}} (\parameter)}$, where we have omitted the dependency on the time $t$. The definition of the Fisher information obtainable from the trajectories is
\begin{equation}
    F^{c, h} \equiv \mathbb{E}_{\text{traj}} \left[ \left( \frac{\partial \log p_{\text{traj}}(\parameter)}{\partial \parameter} \right)^2 \right] \; .
    \label{eq:classical_fisher_information_countinuous}
\end{equation}
We can evaluate the derivative of $ \log p_{\text{traj}}(\parameter)$ as following:
\begin{equation}
\begin{aligned}
    \frac{\partial \log p_{\text{traj}}(\parameter)}{\partial \parameter}
    &= \frac{1}{\braket{\tilde{\psi}_{\text{traj}} (\parameter) | \tilde{\psi}_{\text{traj}} (\parameter)}}
       \Bigl( \braket{\partial_\parameter \tilde{\psi}_{\text{traj}} (\parameter) | \tilde{\psi}_{\text{traj}} (\parameter)} \\
    &\quad + \braket{\tilde{\psi}_{\text{traj}} (\parameter) | \partial_\parameter \tilde{\psi}_{\text{traj}} (\parameter)} \Bigr) \; .
\end{aligned}
\end{equation}
We define $\ket{\partial_\parameter \tilde{\psi}_{\text{traj}} (\parameter)} \equiv \sqrt{p_{\text{traj}}(\parameter)} \ket{\phi_\text{traj} (\parameter)}$, and write the classical Fisher information as
\begin{equation}
\begin{aligned}
    F^{c, h} &= \sum_{\text{traj}} p_{\text{traj}}(\parameter) 
       \bigl( \braket{\psi_{\text{traj}}(\parameter) | \phi_{\text{traj}}(\parameter)} \\
    &\quad + \braket{\phi_{\text{traj}}(\parameter) | \psi_{\text{traj}}(\parameter)} \bigr)^2
\end{aligned}
\label{eq:semi_final_qfi}
\end{equation}
Following~\cite{mattes_designing_2025} we define $A_{\text{traj}} \equiv \braket{\phi_{\text{traj}}(\parameter) | \psi_{\text{traj}}(\parameter)}$, and we write the Fisher information in Eq.~\eqref{eq:semi_final_qfi} under the assumption $A_{\text{traj}} \in \mathbb{R}$, which we will prove to be satisfied in the scenario we consider for photon counting and homodyne detection. With this simplification we can write Eq.~\eqref{eq:semi_final_qfi} as
\begin{equation}
    F^{c, h} = 4 \sum_{\text{traj}} p_{\text{traj}}(\parameter) A_{\text{traj}}^2 \; .
\end{equation}

\textit{Fisher information of the spins}. From Eq.~\eqref{eq:total_qfi_continuous_measurement}, we see that in order to complete the assessment of how much information is retrievable through continuous measurements, we need to compute the quantum Fisher information for the last measurement on the spin ensemble in addition to the Fisher information from the continuous measurements. Although we will always be interested in the asymptotic case where the final state of the ensemble is independent of the trajectory, in this paragraph we work at a generic time $t$. Consider the expectation value of the Fisher information of the spins over the trajectory,
\begin{equation}
\begin{aligned}
    &\mathbb{E}_{\text{traj}}[\mathcal{F}_\parameter(\ket{\psi_\text{traj}(\parameter)})]
    = 4 \sum_{\text{traj}} p_{\text{traj}}(\parameter) \\ & \; \;
       \cdot \left( \braket{\partial_\parameter \psi_{\text{traj}}(\parameter) | \partial_\parameter \psi_{\text{traj}}(\parameter)} + \left( \braket{\partial_\parameter \psi_{\text{traj}}(\parameter) | \psi_{\text{traj}}(\parameter)} \right)^2 \right) \; .
\end{aligned}
\end{equation}
Introducing again the states $\ket{\phi_{\text{traj}}(\parameter)}$, we can obtain an alternative expression for the QFI through the use of the identity
\begin{equation}
    \ket{\partial_\parameter \psi_{\text{traj}}(\parameter)} \equiv 4 \sum_{\text{traj}} \ket{\phi_\text{traj}(\parameter)} - A_{\text{traj}} \ket{\psi_{\text{traj}}(\parameter)} \; ,
\end{equation}
and, assuming $A_{\text{traj}}$ is real, we see that we can rewrite the quantum Fisher information of the system of spins as
\begin{equation}
    \mathcal{F} (\ket{\psi_\text{traj}(\parameter)}) = 4  \left( \braket{\phi_{\text{traj}}(\parameter) | \phi_{\text{traj}}(\parameter)} - A_{\text{traj}}^2 \right) \; ,
\end{equation}
as it has been shown already in Ref.~\onlinecite{mattes_designing_2025}. Since we are in the regime of single parameter estimation, this QFI is achievable through the optimal measurement on the spin ensemble. By putting together the Fisher information of the continuous measurement and the measurement of the spin ensemble (with the expectation value) we get the following expression
\begin{align}
    F^{h, c} &\equiv I^{h, c}(p_\text{traj}) + \mathbb{E}_{\text{traj}} [\mathcal{F}_\parameter(\ket{\psi_\text{traj}(\parameter)})] \nonumber \\ &=
    4 \sum_{\text{traj}} \braket{\tilde{\psi}_{\text{traj}} (\parameter) | \tilde{\psi}_{\text{traj}} (\parameter)} 
    \braket{{\phi}_{\text{traj}} (\parameter) | {\phi}_{\text{traj}} (\parameter)} \; .
    \label{eq:final_expression_calssical_QFI}
\end{align}

\textit{Fisher information of photon counting}. Writing explicitly the expression for $\ket{\phi_{\text{traj}} (\parameter)}$, we easily see that Eq.~\eqref{eq:final_expression_calssical_QFI} is identical to Eq.~\eqref{eq:fisher_photon_counting}. It remains to prove that $A_{\text{traj}}$ is real. This is easily done by considering the numerical expression for the unnormalized evolved state in the Dicke basis. The initial state $|j \rangle$, expressed in this basis, has only real components. The operators $\hat{K}^\parameter_0$ and $\hat{K}^\parameter_1$ in Eq.~\eqref{eq:kraus_photon_counting}, expressed in the Dicke basis, are real matrices for $\parameter=0$, so that also the state $\ket{\psi_{\text{traj}}(\parameter=0)}$ is real. Consider now the derivatives of the Kraus operators, which are necessary to compute $\ket{{\phi}_{\text{traj}} (\parameter)}$:
\begin{equation}
    \frac{\partial \hat{K}^\parameter_0}{\partial \parameter} = -\frac{1}{2} \Gamma \Delta t \frac{\partial (\hat{L}_\parameter^\dagger \hat{L}_\parameter)}{\partial \parameter} \hat{K}^\parameter_0 = -\frac{1}{2} \Gamma \lbrace \hat{J}_x, \hat{J}_z \rbrace \; ,
\end{equation}
and
\begin{equation}
    \frac{\partial \hat{K}^\parameter_1}{\partial \parameter} = \sqrt{\Delta t \Gamma} \frac{\partial \hat{L}_\parameter}{\partial \parameter} = \sqrt{\Delta t \Gamma} \hat{J}_z \; .
\end{equation}
Both these operators, when expressed as matrices in the canonical Dicke basis, have real entries. Therefore, we conclude that $A_{\text{traj}} \equiv \braket{\phi_{\text{traj}}(\parameter) | \psi_{\text{traj}}(\parameter)} \in \mathbb{R}$, and that the Fisher information computed from the trajectories of the photon-counting measurements achieves the quantum Fisher information.

\textit{Fisher information of homodyne detection}. We now repeat the same argument for homodyne detection. Starting from Eq.~\eqref{eq:kraus_homodyne} we see that, with $\hat{H}=0$, the operator $\hat{K}_J$ is real, if we assume $\theta = 0$. When taking the derivative we have that both $\frac{\partial (\hat{L}_\parameter^\dagger \hat{L}_\parameter)}{\partial \parameter}$ and $\frac{\partial \hat{L}_\parameter}{\partial \parameter}$ are real, as proved in the previous paragraph. Therefore, we have that $A_{\text{traj}} \in \mathbb{R}$ also holds for homodyne detection, and this measurement saturates the QFI.

\textit{Limitations of the proof}. This proof, based on the fact that $A_{\text{traj}}$ has no imaginary part, fails when $\hat{H}\neq0$ (which happens when the cavity is detuned) or when the generator of the rotation has imaginary components, as it happens for Eq.~\eqref{eq:initial_state_entangled}, using squeezing in the initial state of the ensemble. This does not necessarily mean that homodyne detection and photon counting cannot achieve the QFI in this case, but it means that the achievability of the QFI though continuous measurement cannot be proven within the framework delineated in this appendix.

\section{Continuous encoding}
\label{sec:continuous_encoding}
In this appendix, we extend the formalism of matched filtering to continuous encoding. We again assess the capabilities of the homodyne and photon counting measurements. We want to sense the parameter $\parameter$ encoded by the Hamiltonian term $\parameter \generator$ acting on the ensemble of spins continuously with the parameter $\parameter$. We indicate with $\mathcal{L}$ the Lindbladian operator describing the superradiant decay of the spin ensemble, so that by adding the continuous encoding we can write $\mathcal{L}_\parameter \equiv \mathcal{L} - i \parameter [ \generator, \cdot]$. We consider the Heisenberg evolved jump operator $J_{-}^\parameter(t)$, and we are interested in evaluating its parametric derivative to plug it into Eq.~\eqref{eq:homodyne_sensitivity}, in order to obtain the homodyne signal. To achieve this the parametric derivative of the evolution operator is useful, which can be computed as the following integral:
\begin{align}
    \left. \frac{\partial}{\partial \parameter}  \, \text{e}^{\mathcal{L}_\parameter t} \right|_{\parameter = 0} 
    &= t \int_0^1 \! \dd s \, \text{e}^{(1-s)\, \mathcal{L} t} \,
    \frac{\partial \mathcal{L}_\parameter}{\partial \parameter} \,
    \text{e}^{s\, \mathcal{L}_\parameter t} \; .
\end{align}
The parametric derivative of the Lindbladian is the commutator, i.e. $\frac{\partial \mathcal{L}_\parameter}{\partial \parameter} = -i [\generator, \cdot]$. Applying this formula to the jump operator gives the final formula for the derivative of the lowering operator $\hat{J}^{\parameter}_{-} (t)$:
\begin{equation}
    \left. \frac{\partial \hat{J}^\parameter_{-} (t)}{\partial \parameter}  \right|_{\parameter = 0} = -i t \int_{0}^{1} \dd s \, \text{e}^{(1-s) \mathcal{L} t} \left( \left[ \generator, \text{e}^{s  \mathcal{L} t} (\hat{J}_{-}) \right] \right) \; .
\end{equation}
From the above expression we can evaluate the homodyne signal
\begin{align}
    & \alpha_I(t) =  - i \sqrt{\Gamma} \left[ e^{i \theta} \left\langle \frac{\partial \hat{J}_{-}^{(\parameter)}(t)}{\partial \parameter} \bigg|_{\parameter = 0} \right\rangle + \text{h.c.} \right] 
    \notag \\
    &= - t \sqrt{\Gamma} \left[ e^{i \theta} \int_{0}^{1} \dd s \, \braket{ \text{e}^{(1-s) \mathcal{L} t} \left( \left[ \generator, \text{e}^{s \mathcal{L} t} (\hat{J}_{-}) \right] \right) } + \text{h.c.} \right] 
    \label{eq:homodyne_sensitivity_countinuous}
\end{align}

The correlation function of the outcomes is still given by Eq.~\eqref{eq:S_spins}, since the infinitesimally small perturbation caused by $\parameter$ doesn't change the state of the system to zeroth order in $\parameter$. The result for the mode maximizing the SNR in Eq.~\eqref{eq:snr_optimizer} is still valid for continuous encoding. In the photon counting scenario, we can derive a similar expression by computing the parametric derivative of the correlation function $g_1(t', t; \parameter)$.

\end{document}